\documentclass[draftcls,onecolumn,12pt]{IEEEtran}
\usepackage[usenames,dvipsnames]{color}
\usepackage{amsfonts}
\usepackage{amsmath}
\usepackage{amssymb}
\usepackage{balance}
\usepackage{bbm}
\usepackage{bm}
\usepackage{cases}
\usepackage{cite}
\usepackage{color}
\usepackage{paralist}
\usepackage{subfigure}
\usepackage[dvips]{graphicx}
\DeclareGraphicsExtensions{.eps}

\usepackage{amsthm}
\newtheorem{thm}{Theorem}

\newtheorem{coro}{Corollary}

\newcommand {\bx}{\mathbf{x}}
\newcommand {\by}{\mathbf{y}}
\newcommand {\bz}{\mathbf{z}}

\newcommand {\bzero}{\mathbf{0}}

\makeatletter
\newcommand{\vast}{\bBigg@{3}}
\makeatother

\begin{document}

\title{Optimizing User Association and Spectrum Allocation in HetNets: A Utility Perspective}

%\author{Yicheng~Lin,~\IEEEmembership{Student~Member,~IEEE,} Wei~Bao,~\IEEEmembership{Student~Member,~IEEE,}
%Wei~Yu,~\IEEEmembership{Fellow,~IEEE} and~Liang~Ben,~\IEEEmembership{Senior~Member,~IEEE}
\author{Yicheng~Lin, Wei~Bao, Wei~Yu, and~Ben~Liang
\thanks{This work was supported by the Natural Sciences and Engineering Research Council (NSERC) of Canada. The materials in this paper have been presented in part at IEEE Global Communications Conference (Globecom), Atlanta, GA, U.S.A., Dec. 2013, and at the 48th Annual Conference on Information Sciences and Systems (CISS), Princeton, NJ, U.S.A., Mar. 2014.

The authors are with The Edward S. Rogers Sr. Department of Electrical and Computer Engineering, University of Toronto, 10 King's College Road, Toronto, Ontario M5S 3G4, Canada (e-mail: \{ylin, wbao, weiyu, liang\}@comm.utoronto.ca).}}

%\markboth{Submitted to IEEE Journal on Selected Areas in Communications} {Lin \MakeLowercase{\textit{et al.}}: Optimizing User Association and Spectrum Allocation in HetNets: A Utility Perspective}

\maketitle

\begin{abstract}
The joint user association and spectrum allocation problem is studied for multi-tier heterogeneous networks (HetNets) in both downlink and uplink in the interference-limited regime. Users are associated with base-stations (BSs) based on the biased downlink received power. Spectrum is either shared or orthogonally partitioned among the tiers. This paper models the placement of BSs in different tiers as spatial point processes and adopts stochastic geometry to derive the theoretical mean proportionally fair utility of the network based on the coverage rate. By formulating and solving the network utility maximization problem, the optimal user association bias factors and spectrum partition ratios are analytically obtained for the multi-tier network. The resulting analysis reveals that the downlink and uplink user associations do not have to be symmetric. For uplink under spectrum sharing, if all tiers have the same target signal-to-interference ratio (SIR), distance-based user association is shown to be optimal under a variety of path loss and power control settings. For both downlink and uplink, under orthogonal spectrum partition, it is shown that the optimal proportion of spectrum allocated to each tier should match the proportion of users associated with that tier. Simulations validate the analytical results. Under typical system parameters, simulation results suggest that spectrum partition performs better for downlink in terms of utility, while spectrum sharing performs better for uplink with power control.
\end{abstract}


%\IEEEpeerreviewmaketitle
\newpage

\section{Introduction}

\begin{figure}[!t]
\centering
\includegraphics[width=3.5in]{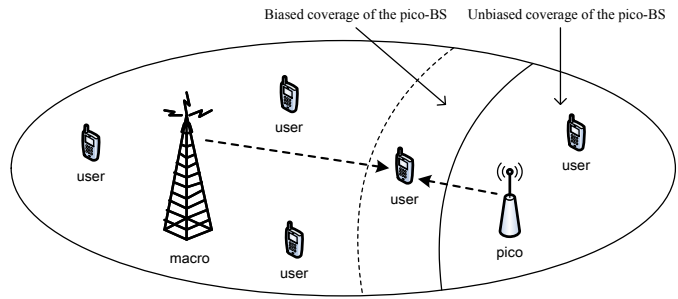}
\caption{An example of a 2-tier HetNet with cell range expansion.} \label{Fig:ScenarioOffloading}
\end{figure}

\IEEEPARstart{F}{uture} wireless networks are expected to accommodate exploding mobile data traffic demands that will severely strain the
traditional single-tier macro cellular access network. Heterogeneous network (HetNet) architecture \cite{AleksandarDamnjanovic/2011/Survey}
provides one possible solution for dealing with this traffic explosion problem. In a HetNet, various types of additional low-power access nodes (e.g., micro, pico, and femto base-stations (BSs)) are deployed to offload macro cell users, forming a multi-tier network overlaid with many small cells (see Fig. \ref{Fig:ScenarioOffloading}). In this architecture, the macro cells offer basic long-range coverage, and the small cells provide short-range but high-quality communication links to nearby users.

The deployment of HetNets, however, also faces many challenges. For example, since network parameters such as transmission power and deployment density are distinct across BS tiers, inter-tier load balancing is a nontrivial issue. Further, as the increased density of small cell transmitters leads to more interference in the network, efficient and practical methods to mitigate interference are critical to network performance.

This paper addresses the joint load balancing (i.e., user association) and spectrum allocation problem in a multi-tier HetNet for both downlink and uplink. Our main insight is that this joint problem can be analytically solved by maximizing a network utility function based on the coverage rate averaged over network spatial topologies and channel realizations.

\subsection{System Modeling Assumptions}

\subsubsection{Spatially Random Deployment}

To account for the irregular deployment of low-power small cell BSs for hot-spot coverage, we assume that BSs of different tiers form independent homogeneous Poisson point processes (PPPs) with different deployment intensities \cite{HanShinJo/2012/Heterogeneous,HarpreetDhillon/2012/Modeling,SarabjotSingh/2013/Offloading}. Users are scattered, and  form an independent PPP. Stochastic geometry \cite{MartinHaenggi/2009/Stochastic,MartinHaenggi/2012/Stochastic} is applied as a basic tool to derive performance metrics in closed forms and to provide system design guidelines.

\subsubsection{Biased User Association}

Associating users to the BSs with the maximum downlink received power may not be the optimal strategy in a HetNet, because in this case most users would tend to connect to high-power macro BSs, thus causing overloading. Although dynamic approaches to user association is possible, this paper adopts the simple and effective \emph{cell range expansion} scheme, also known as \emph{biased user association} \cite{AleksandarDamnjanovic/2011/Survey,PerezLopez/2012/Expanded}, where each BS is assigned a bias factor, and each user is associated with the BS that  provides the maximum received power weighted by its bias. By setting a larger bias towards low-power BSs, traffic can be effectively offloaded to them. (See Fig. \ref{Fig:ScenarioOffloading} as an illustration of a 2-tier network with biased association.) Note that the bias factors are assigned differently across the tiers but are kept the same within a tier, as BSs within a tier are expected to have approximately the same load. The bias should be properly designed such that all users receive adequate quality of service, i.e., it should achieve a tradeoff between signal quality from the users' perspective and load balancing from the BSs' perspective.

\subsubsection{Spectrum Allocation}

This paper considers both spectrum sharing and orthogonal spectrum partition among tiers. Spectrum sharing is more bandwidth efficient, but it exacerbates the inter-tier interference problem, especially when cell range expansion is applied. In the downlink, users offloaded to small cells experience large interference from macro cells; in the uplink with power control, macro users who are far from its associated BS tend to transmit at high power, causing strong interference to small cell BSs. Alternatively, by partitioning the total spectrum into disjoint portions and allocating one partition for each tier (e.g., see \cite{VikramChandrasekhar/2009/Spectrum}), cross-tier interference can be avoided\footnote{In this paper we assume downlink and uplink transmissions in a system are separated via either FDD or TDD, and are treated independently. Hence, there is no spectrum sharing or partition between downlink and uplink.}. This greatly reduces the complexity of interference management especially in an irregular network topology, at the cost of the reduced spectrum usage.

\subsubsection{Utility Optimization}

This paper defines a proportionally fair utility function based on the coverage rate. The mean of such a utility can be derived in a compact and closed form under random deployment of BSs and Rayleigh fading, and can then be used as the objective of an optimization problem. The analytic solution to this network utility maximiztion problem can offer substantial system design insights.

\subsection{Contribution}

This paper aims to optimize the user association and spectrum allocation in a HetNet. The BS powers are assumed to be fixed in the downlink, while fractional power control is used in the uplink. Instead of focusing on coverage probability and rate, as in many previous studies of HetNets using stochastic geometry, we take a network utility maximization approach. In particular, we define the user coverage rate based on a target signal-to-interference ratio (SIR), and derive a closed-form approximation of the mean user proportionally fair utility using stochastic geometry for both downlink and uplink. Such a utility is averaged over BS locations and fading channels, so it does not depend on a specific network realization. An essential part of deriving this utility is to compute the downlink and uplink mean interference for a randomly picked user. By maximizing the utility, we can obtain the optimal bias factor and the optimal proportion of spectrum allocated to each tier. Specifically, we analytically show that
\begin{itemize}
\item Users may choose to associate with different BSs in downlink and uplink for better performance.
\item For uplink under spectrum sharing, if all tiers have the same target SIR, the optimal user association is shown to be distance based under a variety of path loss and power control settings, i.e., each user connects to its nearest BS.
\item For both downlink and uplink, under orthogonal spectrum partition, the optimal proportion of spectrum allocated to each tier is equal to the proportion of users associated with that tier. This suggests a simple and optimal spectrum partition scheme for HetNets.
\end{itemize}
Simulations  validate our analysis. Under  typical system parameters, we observe from simulation that
\begin{itemize}
\item The association bias and spectrum allocation resulted from optimizing the proposed utility match the optimal values obtained via numerical experiments.
\item Orthogonal spectrum partition performs better in terms of utility for downlink systems, while spectrum sharing performs better for uplink systems with power control.
\end{itemize}

\subsection{Related Work}
The use of spatial random point processes to model transmitters and receivers in wireless networks has been considered extensively in the literature. It allows tools from stochastic geometry \cite{MartinHaenggi/2009/Stochastic,MartinHaenggi/2012/Stochastic} to be used to characterize performance metrics analytically. For example, the random network topology is assumed in characterizing the coverage and rate \cite{JeffreyAndrews/2011/Tractable} as well as handover \cite{ThanTungVu/2014/Analytical} in traditional cellular networks. Stochastic geometry based analysis can also be extended to multi-tier HetNets: the flexible user association among different tiers is analyzed in \cite{HanShinJo/2012/Heterogeneous}, where the coverage and rate are analyzed; open-access and closed-access user association are discussed in \cite{HarpreetDhillon/2012/Modeling}; the distribution of the per-user rate is derived in \cite{SarabjotSingh/2013/Offloading} by considering the cell size and user distribution in the random networks. However, none of these works characterizes user performance from a network utility perspective, which models the tradeoff between rate and fairness.

For the user association problem, one of the prior approaches in the literature involves heuristic greedy search, i.e., adding users that improve a certain metric to the BS in a greedy fashion, as in \cite{KyuhoSon/2009/Dynamic} and \cite{RiteshMadan/2010/Cell} for single-tier networks and multi-tier HetNets, respectively. Another prior approach involves a utility maximization framework and  pricing-based association methods, see \cite{JangWonLee/2006/Joint} for single-tier networks and \cite{QiaoyangYe/2013/User,KaimingShen/2014/Distributed} for HetNets. In \cite{MingyiHong/2012/Mechanism,MingyiHong/2013/Joint}, the association problem is jointly considered with resource allocation using the game theoretical approach. These solutions are dynamic and require real-time computations based on channel and topology realization. The cell range expansion scheme \cite{AleksandarDamnjanovic/2011/Survey,PerezLopez/2012/Expanded} considered in this paper is semi-static and simple to implement. However, the bias factors are usually empirically determined through system-level performance evaluation \cite{AleksandarDamnjanovic/2011/Survey}. The effect of biased offloading has been investigated for multi-tier HetNets in \cite{HanShinJo/2012/Heterogeneous,SarabjotSingh/2013/Offloading} under random topology, where the optimal bias in terms of SIR and rate coverage is determined through numerical evaluation. In our work, the optimal bias factor of each tier are derived through analytical network utility optimization.

For the spectrum allocation problem, disjoint spectrum partition between macro and femto tiers has been considered in prior works. The authors in \cite{VikramChandrasekhar/2009/Spectrum} analytically determine the spectrum partition between the two tiers that maximizes the network-wide area spectral efficiency. Stochastic geometry is used in \cite{WangChiCheung/2012/Throughput} to study the optimal spectrum partition by formulating the throughput maximization problem subject to constraints on coverage probabilities. Biased user association and spectrum partition can be jointly considered. The authors of \cite{SarabjotSingh/2014/Joint} analyze the rate coverage for a two-tier topology and provide trends with respect to the spectrum partition fraction. However, no optimal partition is analytically given. For a general multi-tier network, spectrum partition and user association are optimized analytically in the downlink in terms of the user rate in \cite{WeiBao/2014/Structured} and rate coverage in \cite{SanamSadr/2014/Tier}. Different from these works, under the orthogonal spectrum allocation assumption, we analytically determine the optimal inter-tier spectrum partition in terms of the mean user utility for both downlink and uplink.

Most of the previous works on HetNets focus on the downlink. A key difference in uplink as compared to downlink is that fractional power control is often used in uplink to fully or partially compensate for the path loss, e.g., as defined in 3GPP-LTE \cite{Ericsson/2007/R1-074850}. The influence of fractional power control on system performance is studied in various works, e.g., \cite{CarlosUbedaCastellanos/2008/Performance,ArneSimonsson/2008/Uplink,Mullner/2009/Contrasting} under regular hexagonal topology. For networks with random topology and accounting for fractional power control, \cite{ThomasNovlan/2013/Analytical} analytically derives uplink SIR and rate distribution for single-tier networks; \cite{VikramChandrasekhar/2009/Uplink} investigates uplink outage capacity for two-tier networks with shared spectrum; \cite{HeshamElsawy/2014/Stochastic} extends the analysis to multi-tier uplink networks in terms of outage probability and spectral efficiency. In this paper, the mean user utility of random multi-tier HetNets in uplink with fractional power control is analyzed and optimized.

Part of this work has appeared in \cite{YichengLin/2013/Optimizing,YichengLin/2014/Joint}, which contain the analysis and optimization of the downlink case.

\subsection{Organization}

Section II presents the system model. Section III derives the proposed mean user utility in both downlink and uplink. We present the optimization results of the utility function over the user association and spectrum allocation in Section IV for both downlink and uplink. Section V validates the results through numerical simulation and Section VI concludes the paper.

\section{System Model}

\subsection{Multi-Tier Network Topology}

Following conventional stochastic modeling of HetNets \cite{HanShinJo/2012/Heterogeneous,HarpreetDhillon/2012/Modeling,SarabjotSingh/2013/Offloading}, we consider a total of $K$ tiers of BSs. BSs in the $k$-th tier ($1 \le k \le K$) (or tier-$k$ BSs) are modeled as an independent homogeneous PPP $\Phi_{k}=\{\bx_{k,1},\bx_{k,2},\ldots\}$ with intensity $\lambda_k$ on two dimensional plane, where $\bx_{k,i}$ is the location of the $i$-th BS in the $k$-th tier (or BS $(k,i)$ for simplicity). Without loss of generality, we assume $\lambda_1 < \lambda_2 < \ldots < \lambda_K$. The superposition of all BS tiers is denoted as $\Phi = \bigcup_k \Phi_k$. The multi-tier PPP cell topology forms a multiplicatively weighted Poisson Voronoi tessellation \cite{AtsuyukiOkabe/2009/Spatial}, as in Fig. \ref{Fig:PPPTopology2Tiers}.

Users form another independent homogeneous PPP $\Psi=\{\by_1,\by_2,\ldots\}$ with intensity $\lambda_u$, where $\by_i$ is the $i$-th user's location. In the uplink, for a given spectrum resource block, the interference comes from the users scheduled by other BSs on the same resource block. Only one user out of all users associated with each cell is scheduled to transmit on each spectrum resource block. We let $\Psi' \subset \Psi$ be such a scheduled user set over the entire network on an arbitrarily chosen resource block, and further partition $\Psi' = \bigcup_k \Psi_k$ where $\Psi_k = \{\by_{k,1},\by_{k,2},\ldots\}$ contains users associated with and scheduled by tier-$k$ BSs, and $\by_{k,i}$ is the location of the user scheduled by BS $(k,i)$ (or user $(k,i)$ for simplicity). Unlike $\Phi_k$ for BSs, the user point process $\Psi_k$ does not form a PPP.

\subsection{Path Loss and Power Model}

Suppose that $P^{(t)}(\bz_1)$ is the transmit power from a BS or user at location $\bz_1$, the received power at location $\bz_2$ is modeled as $P^{(r)}(\bz_2) = P^{(t)}(\bz_1) \left|\bz_1-\bz_2\right|^{-\alpha} g(\bz_1,\bz_2)$, where $\left|\bz_1-\bz_2\right|^{-\alpha}$ is the propagation loss with a path loss exponent $\alpha$ ($\alpha>2$), and $g(\bz_1,\bz_2)$ is the small-scale channel power fading between $\bz_1$ and $\bz_2$. We assume that $\alpha$ is a constant for all tiers\footnote{Systems aggregating multiple component carriers may have multiple different path loss exponents \cite{XingqinLin/2013/Modeling}, which is beyond the scope of this paper.}. To model Rayleigh fading, we assume that $g(\bz_1,\bz_2)$ is independently and identically distributed with an exponential distribution of unit mean. Shadowing is ignored here for simplicity and tractability. Note that the randomness of the node locations can approximately model shadowing: as shadowing variance increases, the resulting propagation losses between the BSs and the typical user in a grid network converge to those in a Poisson distributed network \cite{BartlomiejBlaszczyszyn/2013/Using}.

In the downlink, the transmit power of tier-$k$ BSs is $P_k$. In the uplink we consider fractional power control \cite{Ericsson/2007/R1-074850,ThomasNovlan/2013/Analytical}. Let the user power before doing power control be $P_u$. Suppose that a user located at $\by$ is associated with a BS located at $\bx$, the uplink transmit power after power control is $P_u\left|\by-\bx\right|^{\epsilon \alpha}$, where $\epsilon \in \left[ 0,1 \right]$ is the power control factor. The received power at this BS is thus $P_u \left|\by-\bx\right|^{(\epsilon - 1) \alpha}g(\by,\bx)$. Note that full power control is achieved at $\epsilon = 1$, where the BS received power is a constant irrespective of the distance between the user and the BS. No power control is applied when $\epsilon = 0$.

\subsection{Biased User Association}

\begin{figure}[!t]
\centering
\subfigure[Without Biasing]{\includegraphics[width=1.67in]{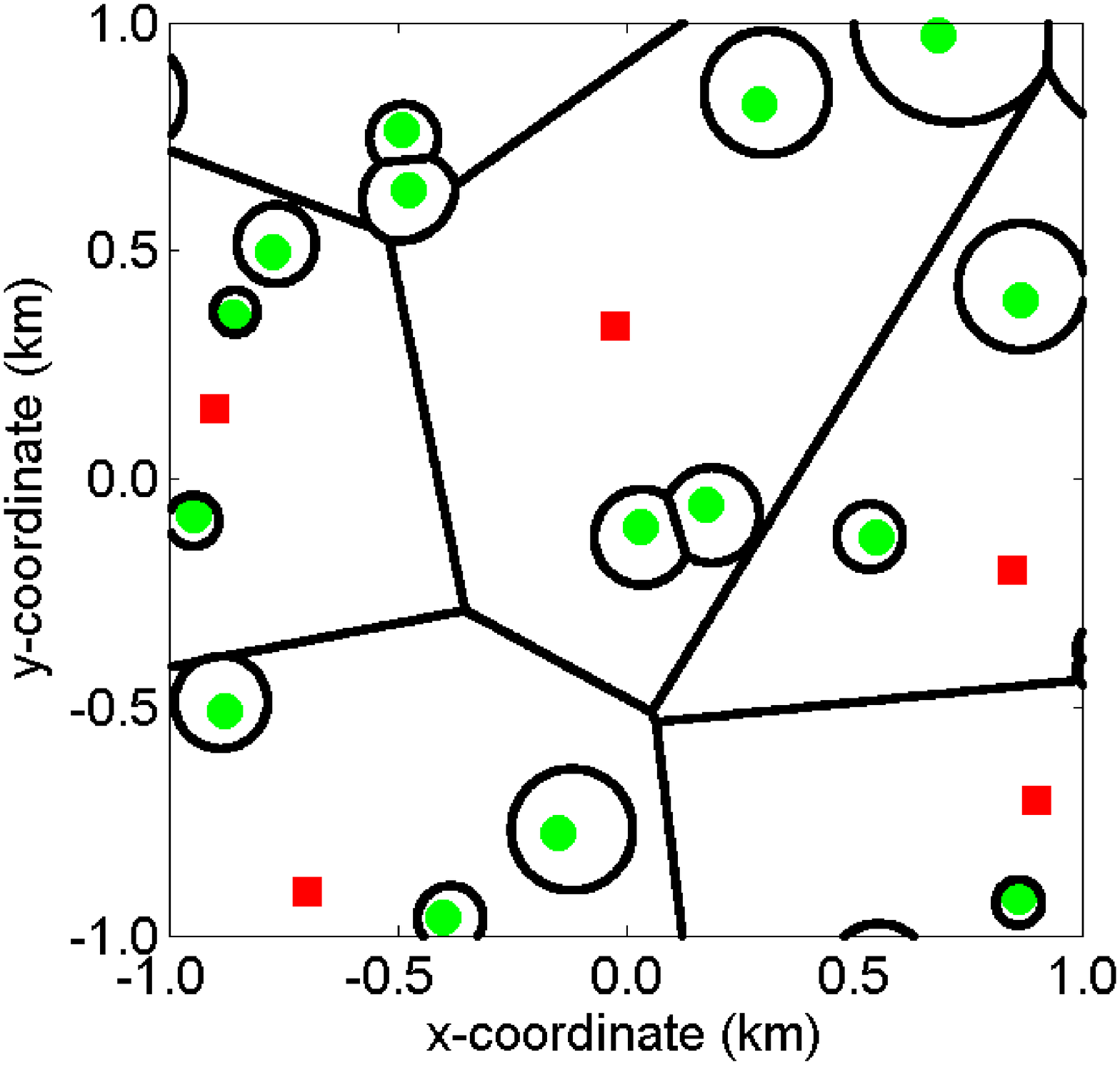}
\label{Fig:PPPTopology2TiersNoBias}}
\subfigure[With Biasing]{\includegraphics[width=1.67in]{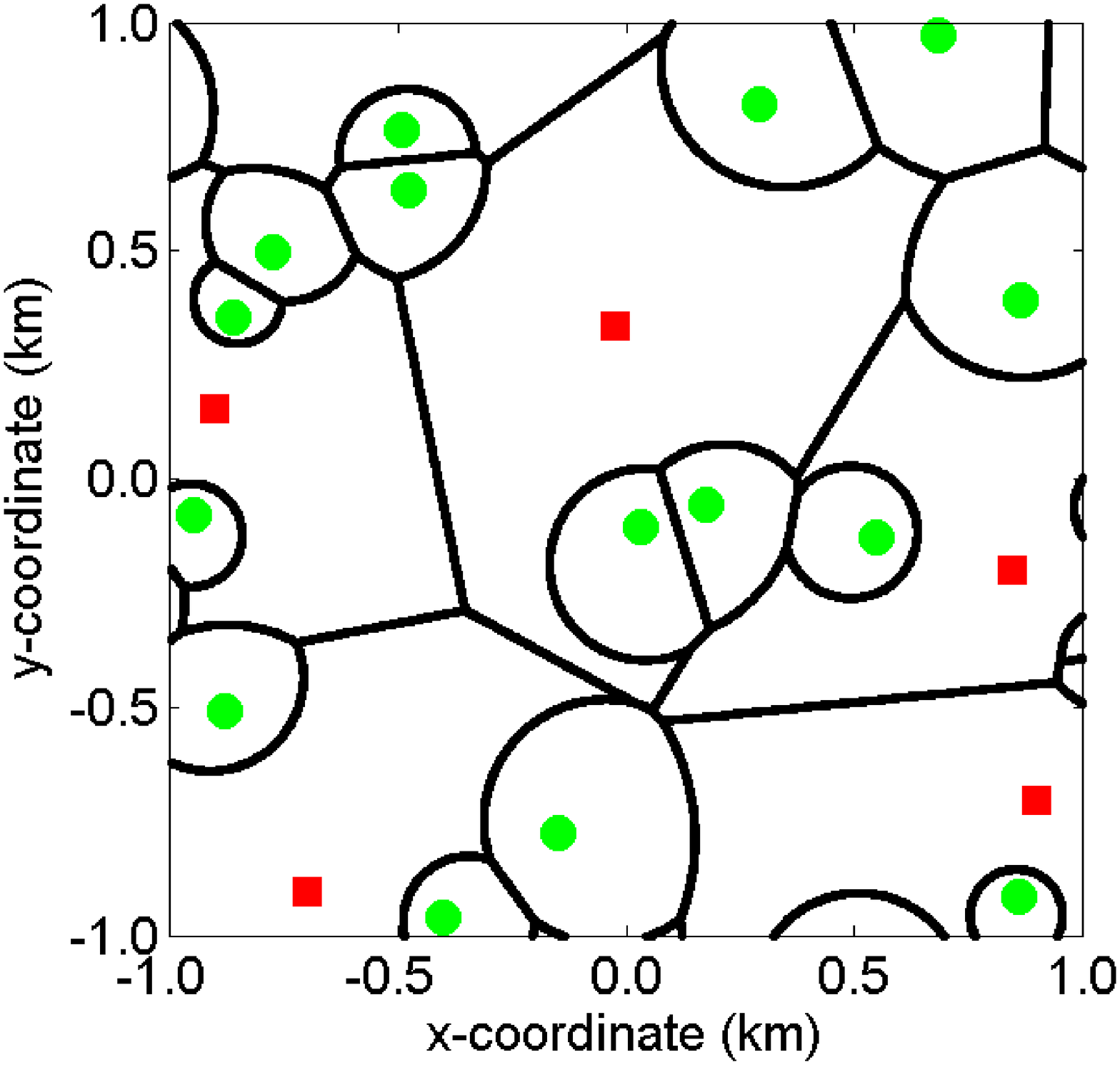}
\label{Fig:PPPTopology2TiersWithBias}} \caption{Two-tier cell topology with and without biasing.} \label{Fig:PPPTopology2Tiers}
\end{figure}

User association is determined by the downlink received power from BSs measured at the user side. A user located at $\by$ is associated with a tier-$k$ BS if it provides the maximum biased received power \cite{HanShinJo/2012/Heterogeneous}
\begin{equation} \label{Eq:AssociationRule}
P_k B_k \hspace{-0.5mm} \left( \hspace{-0.5mm} \min_i \left|\bx_{k,i}-\by \right| \hspace{-0.5mm} \right)^{\hspace{-1mm} -\alpha} \hspace{-0.5mm} \ge \hspace{-0.5mm} P_j B_j \hspace{-0.5mm} \left( \hspace{-0.5mm} \min_{i'} \left|\bx_{j,i'}-\by \right| \hspace{-0.5mm} \right)^{\hspace{-1mm} -\alpha}, \forall j
\end{equation}
where $B_k$ is the bias indicating the connecting preference of a user toward tier-$k$ BSs. An example of the coverage of a two-tier network with and without biasing is shown in Fig. \ref{Fig:PPPTopology2Tiers}.

Using (\ref{Eq:AssociationRule}), the probability of a user being associated with a tier-$k$ BS, denoted as $A_k$, can be derived as in \cite{HanShinJo/2012/Heterogeneous}
\begin{equation} \label{Eq:AssociationProbability}
A_k \hspace{-0.3mm} = \hspace{-0.3mm} \frac{ \lambda_k \left( P_k B_k \right)^{2/\alpha} }{ \sum_{j=1}^K \lambda_j \left( P_j B_j \right)^{2/\alpha} } \hspace{-0.3mm} = \hspace{-0.3mm} \left[ \sum_{j=1}^{K} { \hat{\lambda}_{jk} \left( \hat{P}_{jk} \hat{B}_{jk} \right)^{2 / \alpha} } \right]^{-1} \hspace{-3mm},
\end{equation}
where $\hat{\lambda}_{jk} \triangleq \frac{ \lambda_j }{ \lambda_k }$, $\hat{P}_{jk} \triangleq \frac{ P_j }{ P_k }$, $\hat{B}_{jk} \triangleq \frac{ B_j }{ B_k }$.

We can also derive the corresponding bias $\left\{ B_k \right\}_{\forall k}$ from values of $\left\{ A_k \right\}_{\forall k}$. Since the effects of $\left\{ B_k \right\}_{\forall k}$ remain the same if a positive constant is multiplied to all of them, without loss of generality, we can assume the bias of tier-$K$ BSs (with the largest deployment intensity) to be one, i.e., setting $B_K = 1$, and recover $\left\{ B_k \right\}_{\forall k}$ from $\left\{ A_k \right\}_{\forall k}$ via simple manipulation:
\begin{equation} \label{Eq:BasA}
B_k \leftarrow \left( \hat{\lambda}_{Kk} \hat{A}_{Kk}^{-1} \right)^{\alpha/2} \hat{P}_{Kk},
\end{equation}
where $\hat{A}_{jk} \triangleq \frac{A_j}{A_k}$.\footnote{By convention, the maximum bias of a tier is 1 ($0$dB). If any other tier has a larger bias than tier-$K$ after computing (\ref{Eq:BasA}), we need to add a normalization step for all tiers $B_k \leftarrow B_k / \max_j \left( B_j \right)$.}

\subsection{Spectrum Allocation among Tiers}

The total system spectrum is denoted as $W$. The spectrum allocated to tier-$k$ is denoted as $W_k$. Two types of spectrum allocation schemes are considered for the multi-tier topology in this paper: the orthogonal partition scheme where each BS tier is allocated non-overlapping spectrum, i.e., $W_k = \eta_k W$ and $\sum_k \eta_k = 1$, and the full reuse scheme where all BS tiers in the network share the all spectrum band, i.e., $W_k = W, \forall k$.

\subsection{Coverage Probability and User Coverage Rate}

In this paper, we assume that the background noise is negligible and the system is interference limited\footnote{With a noise term, the mean utility can still be derived and optimized numerically, but the optimization results cannot be obtained in closed forms.}, which is a valid assumption for a dense network. The coverage probability of a user (either in the downlink or uplink) is defined as \cite{JeffreyAndrews/2011/Tractable}
\begin{equation}
C_k = \mathbb{P} \left( \text{SIR}_k > \tau_k \right),
\end{equation}
where $\tau_k$ is the target SIR of this user given that it is associated with tier-$k$ BSs.

Further, on each spectrum resource block, we assume that the user does not obtain a positive rate if the SIR is below $\tau_k$, and is served with a constant rate otherwise. The spectrum efficiency of the user in tier-$k$ under this model hence has a binary form (in nats/s/Hz) \cite{WangChiCheung/2012/Throughput,HarpreetDhillon/2013/DownlinkMIMO}
\begin{equation} \label{Eq:SE}
r_k = \log \left( 1 + \tau_k \right) \bm{1} \left( \text{SIR}_k > \tau_k \right),
\end{equation}
where $\bm{1} \left( \cdot \right)$ is the indicator function. This model corresponds to a transmission scheme with a fixed modulation and coding format, but results in a closed-form utility expression amenable to optimization. The user rate is obtained by summing these spectral efficiencies across the spectrum $\beta_k$ allocated to this user. The mean of this user rate can be computed as
\begin{equation} \label{Eq:MeanRate}
R_k = \beta_k \mathbb{E} \left( r_k \right) = \beta_k C_k \log \left( 1 + \tau_k \right).
\end{equation}

\section{Mean User Utility}

In this section, we study the mean utility of a randomly chosen user (termed the typical user), communicating with its serving BS (termed the typical BS). We are interested in the typical user since its mean performance represents the mean system performance.

The mean utility of the typical user is
\begin{equation} \label{Eq:MeanPerUserUtility}
U = \sum_{k=1}^K A_k U_k,
\end{equation}
where $U_k$ is the mean user utility given that the typical user is associated with a tier-$k$ BS. In this paper, we adopt a new notion of the coverage-rate-based proportionally fair utility, defined as the logarithm of the user coverage rate:
%\begin{align} \label{Eq:UtilityPerTier}
%U_k &= \mathbb{E} \left[ \log \left( R_k \right) \right] \nonumber \\
%&= \log\left[ \log \left( 1 + \tau_k \right) \right] + \mathbb{E} \left[ \log \left( \beta_k \right) \right] + \mathbb{E} \left[ \log \left( C_k \right)\right].
%\end{align}
\begin{equation} \label{Eq:UtilityPerTier}
U_k = \mathbb{E} \left[ \log \left( R_k \right) \right] = \log\left[ \log \left( 1 + \tau_k \right) \right] + \mathbb{E} \left[ \log \left( \beta_k \right) \right] + \mathbb{E} \left[ \log \left( C_k \right)\right].
\end{equation}
The proportionally fair utility \cite{DavidTse/1999/Transmitter,FrancoisBaccelli/2014/Analysis} captures a tradeoff between opportunism and user fairness, by encouraging low-rate users to improve their rates while saturating the utility gain of high-rate users. The use of the logarithmic function also separates the computation of $\beta_k$ and $C_k$, although they are not statistically independent. Note that the utility of each user is based on its mean rate averaged over the fading channel; while the mean system utility is the average of such utilities of all users over the network topology, which is equivalent to the mean utility of the typical user.

The mean user utility therefore relies on the mean logarithm of the user spectrum $\mathbb{E} \left[ \log \left( \beta_k \right) \right]$ and the mean logarithm of coverage probability $\mathbb{E} \left[ \log \left( C_k \right) \right]$. In the following, we give an upper bound for $\mathbb{E} \left[ \log \left( \beta_k \right) \right]$ and analytically derive $\mathbb{E} \left[ \log \left( C_k \right) \right]$ for both downlink and uplink.

\subsection{Mean Logarithm of Per-User Spectrum}

To compute the mean logarithm of per-user spectrum $\mathbb{E} \left[ \log \left( \beta_k \right) \right]$, we assume that all the associated users of a particular BS are allocated equal amount of spectrum bandwidth. Such an equal inter-user spectrum allocation is widely adopted in the research literature (e.g., \cite{HanShinJo/2012/Heterogeneous,QiaoyangYe/2013/User,WangChiCheung/2012/Throughput}). Equal allocation can be shown to maximize the proportionally fair utility under fixed channel and interference pattern \cite{QiaoyangYe/2013/User}. It is also easily implemented by round-robin scheduling.

Even with equal spectrum allocation, $\mathbb{E} \left[ \log \left( \beta_k \right) \right]$ of the typical user is still difficult to compute exactly. We resort to its upper bound using the concavity of the logarithmic function
\begin{equation} \label{Eq:PerUserResource}
\mathbb{E} \left[ \log (\beta_k) \right] \le \log \left[ \mathbb{E} (\beta_k) \right] \overset{(a)}{\lesssim} \log \left( \frac{W_k \lambda_k}{A_k \lambda_u} \right),
\end{equation}
The proof of the upper bound (a) is shown in Appendix A, where the bound is also shown to be tight when the user spatial intensity is much larger than the BS deployment intensity. An intuitive interpretation of (\ref{Eq:PerUserResource}) is that the mean user spectrum $\beta_k$ is the ratio of the total spectrum $W_k$ to the average user number per BS in the $k$-th tier, where the latter is approximately the ratio of the user spatial intensity $A_k \lambda_u$ to the BS deployment intensity $\lambda_k$ of the $k$-th tier. A similar approximation is adopted in \cite{HanShinJo/2012/Heterogeneous}.

\subsection{Mean Logarithm of Coverage Probability in the Downlink}

We derive the mean logarithm of coverage probability of the typical user in the downlink, assuming that the typical BS is in the $k$-th tier. Let $(k,0)$ indicate the typical BS and the typical user. Due to the stationarity of BS and user point processes, we define the coordinates so that the typical user is located at $\by_{k,0} = \bzero$, and consequently $\left|\bx_{k,0}\right| = \min_i \left|\bx_{k,i}\right|$. Both the spectrum sharing and orthogonal spectrum partition schemes are considered.

\subsubsection{Spectrum Sharing}

The SIR of the typical user associated with tier-$k$ BSs is
\begin{equation}
\text{SIR}_k = \frac{ P_k \left|\bx_{k,0}\right|^{-\alpha} g\left(\bx_{k,0},\bzero\right) }{ I_{\Phi} },
\end{equation}
where $I_{\Phi}$ denotes the interference from all tiers of BSs
\begin{equation}
I_{\Phi} = \sum_{j=1}^{K} { \sum_{ i: \bx_{j,i} \in \Phi_j \backslash \bx_{k,0} } { P_j \left|\bx_{j,i}\right|^{-\alpha} g(\bx_{j,i},\bzero) } }.
\end{equation}
The interference is summed over the PPP $\Phi_j$ of each BS tier $j$, excluding the serving BS. From (\ref{Eq:AssociationRule}), given that the typical user is associated with the $k$-th BS tier, the length of interfering links and that of the serving link has the following relationship
\begin{equation} \label{Eq:DownlinkLengthConstraint}
\left|\bx_{j,i}\right| > \left|\bx_{k,0}\right| \left( \hat{P}_{jk} \hat{B}_{jk} \right)^{1 / \alpha}, \; \forall (j,i) \neq (k,0).
\end{equation}

The mean logarithm of the coverage probability is averaged over all the possible typical user locations (in terms of the serving link length $\left|\bx_{k,0}\right|$) and the interference from randomly located BSs. Since from (\ref{Eq:DownlinkLengthConstraint}) the interference depends on $\left|\bx_{k,0}\right|$, we first average over the interference for each value of $\left|\bx_{k,0}\right|$, then average over the distribution of $\left|\bx_{k,0}\right|$
%\begin{align} \label{Eq:ELogCDownlinkShared1}
%&\mathbb{E}_{\left|\bx_{k,0}\right|,\Phi,\mathbf{g}} \left[ \log \left( C_k \right)\right] = \mathbb{E}_{\left|\bx_{k,0}\right|} \left\{ \mathbb{E}_{\Phi,\mathbf{g}} \left[ \log \left( C_k \right) \big| \left|\bx_{k,0}\right| \right] \right\} \nonumber \\
%=& \int_0^\infty \mathbb{E}_{\Phi,\mathbf{g}} \left[ \log \left( C_k \right) \big| \left|\bx_{k,0}\right|=r \right] f_{\left|\bx_{k,0}\right|}(r) \text{d}r,
%\end{align}
\begin{equation} \label{Eq:ELogCDownlinkShared1}
\mathbb{E}_{\left|\bx_{k,0}\right|,\Phi,\mathbf{g}} \left[ \log \left( C_k \right)\right] = \mathbb{E}_{\left|\bx_{k,0}\right|} \left\{ \mathbb{E}_{\Phi,\mathbf{g}} \left[ \log \left( C_k \right) \big| \left|\bx_{k,0}\right| \right] \right\} = \int_0^\infty \mathbb{E}_{\Phi,\mathbf{g}} \left[ \log \left( C_k \right) \big| \left|\bx_{k,0}\right|=r \right] f_{\left|\bx_{k,0}\right|}(r) \text{d}r,
\end{equation}
where the probability density function (PDF) of the distance $\left|\bx_{k,0}\right|$ between the typical user and its serving BS, under the biased user association, is given in \cite{HanShinJo/2012/Heterogeneous} as
%\begin{align} \label{Eq:DistanceDistribution}
%f_{\left|\bx_{k,0}\right|}(r) &= 2\pi r \frac{\lambda_k}{A_k} \exp\left[ - \pi r^2 \sum_{j=1}^{K} \lambda_j \left( \hat{P}_{jk} \hat{B}_{jk} \right)^{2/\alpha} \right] \nonumber \\
%&\overset{(a)}{=} 2\pi r \frac{\lambda_k}{A_k} \exp \left( - \pi r^2 \frac{\lambda_k}{A_k} \right), \quad r \ge 0,
%\end{align}
\begin{equation} \label{Eq:DistanceDistribution}
f_{\left|\bx_{k,0}\right|}(r) = 2\pi r \frac{\lambda_k}{A_k} \exp\left[ - \pi r^2 \sum_{j=1}^{K} \lambda_j \left( \hat{P}_{jk} \hat{B}_{jk} \right)^{2/\alpha} \right] \overset{(a)}{=} 2\pi r \frac{\lambda_k}{A_k} \exp \left( - \pi r^2 \frac{\lambda_k}{A_k} \right), \quad r \ge 0,
\end{equation}
where (a) is obtained by using (\ref{Eq:AssociationProbability}).

Conditioned on the serving link length $\left|\bx_{k,0}\right|$, we have
%\begin{align} \label{Eq:ELogCDownlinkShared2}
%&\mathbb{E}_{\Phi,\mathbf{g}} \left[ \log \left( C_k \right) \big| \left|\bx_{k,0}\right|=r \right] \nonumber \\
%=& \mathbb{E}_{\Phi,\mathbf{g}} \left\{ \log \left[ \mathbb{P} \left( \text{SIR}_k > \tau_k \right) \right] \big| \left|\bx_{k,0}\right|=r \right\} \nonumber \\
%=& \mathbb{E}_{\Phi,\mathbf{g}} \left\{ \log \left[ \mathbb{P} \left( g\left(\bx_{k,0},\bzero\right) > \tau_k P_k^{-1} r^\alpha I_\Phi \right) \right] \big| \left|\bx_{k,0}\right|=r \right\} \nonumber \\
%=& \mathbb{E}_{\Phi,\mathbf{g}} \left\{ \log \left[ \exp \left( -\tau_k P_k^{-1} r^\alpha I_\Phi \right) \right] \big| \left|\bx_{k,0}\right|=r \right\} \nonumber \\
%=& -\tau_k P_k^{-1} r^\alpha \mathbb{E}_{\Phi,\mathbf{g}} \left( I_\Phi \big| \left|\bx_{k,0}\right|=r \right),
%\end{align}
\begin{align} \label{Eq:ELogCDownlinkShared2}
\mathbb{E}_{\Phi,\mathbf{g}} \left[ \log \left( C_k \right) \big| \left|\bx_{k,0}\right|=r \right] &= \mathbb{E}_{\Phi,\mathbf{g}} \left\{ \log \left[ \mathbb{P} \left( \text{SIR}_k > \tau_k \right) \right] \big| \left|\bx_{k,0}\right|=r \right\} \nonumber \\
&= \mathbb{E}_{\Phi,\mathbf{g}} \left\{ \log \left[ \mathbb{P} \left( g\left(\bx_{k,0},\bzero\right) > \tau_k P_k^{-1} r^\alpha I_\Phi \right) \right] \big| \left|\bx_{k,0}\right|=r \right\} \nonumber \\
&= \mathbb{E}_{\Phi,\mathbf{g}} \left\{ \log \left[ \exp \left( -\tau_k P_k^{-1} r^\alpha I_\Phi \right) \right] \big| \left|\bx_{k,0}\right|=r \right\} \nonumber \\
&= -\tau_k P_k^{-1} r^\alpha \mathbb{E}_{\Phi,\mathbf{g}} \left( I_\Phi \big| \left|\bx_{k,0}\right|=r \right),
\end{align}
where we have assumed $g\left(\bx_{k,0},\bzero\right) \sim \exp(1)$. The downlink mean interference is
%\begin{align} \label{Eq:MeanInterferenceDownlinkShared}
%&\mathbb{E}_{\Phi,\mathbf{g}} \left( I_\Phi \big| \left|\bx_{k,0}\right|=r \right) \nonumber \\
%=& \sum_{j=1}^{K} { P_j \mathbb{E}_{\Phi_j} \left( \sum_{i: \bx_{j,i} \in \Phi_j \backslash \bx_{k,0} } \left|\bx_{j,i}\right|^{-\alpha} \Bigg| \left|\bx_{k,0}\right|=r \right) } \nonumber \\
%=& \sum_{j=1}^{K} { P_j \mathbb{E}_{\Phi_j} \hspace{-1mm} \left[ \sum_{i: \bx_{j,i} \in \Phi_j } \hspace{-2mm} \left|\bx_{j,i}\right|^{-\alpha} \mathbf{1} \hspace{-0.5mm} \left( \bx_{j,i} \neq \bx_{k,0} \right) \hspace{-0.5mm} \Bigg| \left|\bx_{k,0}\right|=r \right] } \nonumber \\
%\overset{(a)}{=}& \sum_{j=1}^{K} P_j \mathbb{E}_{\Phi_j} \hspace{-1mm} \left\{ \hspace{-0.5mm} \sum_{i: \bx_{j,i} \in \Phi_j } \hspace{-2mm} \left|\bx_{j,i}\right|^{-\alpha} \mathbf{1} \hspace{-0.5mm} \left[ \left|\bx_{j,i}\right| > r \left( \hat{P}_{jk} \hat{B}_{jk} \right)^{\hspace{-0.5mm} 1/\alpha} \right] \hspace{-0.5mm} \right\} \nonumber \\
%\overset{(b)}{=}& \sum_{j=1}^{K} { P_j 2\pi \lambda_j \int_{ r \left( \hat{P}_{jk} \hat{B}_{jk} \right)^{1 / \alpha} }^{\infty} x^{-\alpha} x \text{d}x } \nonumber \\
%=& \frac{2 \pi }{\alpha-2} r^{2-\alpha} \sum_{j=1}^{K} { P_j \lambda_j \left( \hat{P}_{jk} \hat{B}_{jk} \right)^{ 2 / \alpha - 1 } },
%\end{align}
\begin{align} \label{Eq:MeanInterferenceDownlinkShared}
\mathbb{E}_{\Phi,\mathbf{g}} \left( I_\Phi \big| \left|\bx_{k,0}\right|=r \right) &= \sum_{j=1}^{K} { P_j \mathbb{E}_{\Phi_j} \left( \sum_{i: \bx_{j,i} \in \Phi_j \backslash \bx_{k,0} } \left|\bx_{j,i}\right|^{-\alpha} \Bigg| \left|\bx_{k,0}\right|=r \right) } \nonumber \\
&= \sum_{j=1}^{K} { P_j \mathbb{E}_{\Phi_j} \hspace{-1mm} \left[ \sum_{i: \bx_{j,i} \in \Phi_j } \hspace{-2mm} \left|\bx_{j,i}\right|^{-\alpha} \mathbf{1} \hspace{-0.5mm} \left( \bx_{j,i} \neq \bx_{k,0} \right) \hspace{-0.5mm} \Bigg| \left|\bx_{k,0}\right|=r \right] } \nonumber \\
&\overset{(a)}{=} \sum_{j=1}^{K} P_j \mathbb{E}_{\Phi_j} \hspace{-1mm} \left\{ \hspace{-0.5mm} \sum_{i: \bx_{j,i} \in \Phi_j } \hspace{-2mm} \left|\bx_{j,i}\right|^{-\alpha} \mathbf{1} \hspace{-0.5mm} \left[ \left|\bx_{j,i}\right| > r \left( \hat{P}_{jk} \hat{B}_{jk} \right)^{\hspace{-0.5mm} 1/\alpha} \right] \hspace{-0.5mm} \right\} \nonumber \\
&\overset{(b)}{=} \sum_{j=1}^{K} { P_j 2\pi \lambda_j \int_{ r \left( \hat{P}_{jk} \hat{B}_{jk} \right)^{1 / \alpha} }^{\infty} x^{-\alpha} x \text{d}x } \nonumber \\
&= \frac{2 \pi }{\alpha-2} r^{2-\alpha} \sum_{j=1}^{K} { P_j \lambda_j \left( \hat{P}_{jk} \hat{B}_{jk} \right)^{ 2 / \alpha - 1 } },
\end{align}
where $\mathbf{1}(\cdot)$ is the indicator function, and (a) is from the inequality (\ref{Eq:DownlinkLengthConstraint}). Campbell's Formula \cite{MartinHaenggi/2012/Stochastic} is used in (b) with polar coordinates.

Substituting (\ref{Eq:MeanInterferenceDownlinkShared}) into (\ref{Eq:ELogCDownlinkShared2}) we have
\begin{equation} \label{Eq:ELogCDownlinkShared3}
\mathbb{E}_{\Phi,\mathbf{g}} \left[ \log \left( C_k \right) \big| \left|\bx_{k,0}\right|=r \right] = \frac{-2 \pi \tau_k}{\alpha-2} r^2 \sum_{j=1}^{K} { \lambda_j \hat{P}_{jk}^{2 / \alpha} \hat{B}_{jk}^{ 2 / \alpha - 1 } }.
\end{equation}

Now deconditioning with respect to $\left|\bx_{k,0}\right|$ in (\ref{Eq:ELogCDownlinkShared3}) using (\ref{Eq:DistanceDistribution}), and after some manipulation, (\ref{Eq:ELogCDownlinkShared1}) becomes
%\begin{align} \label{Eq:ELogCDownlinkShared4}
%\mathbb{E}_{\left|\bx_{k,0}\right|,\Phi,\mathbf{g}} \left[ \log \left( C_k \right) \right] &= \frac{-2 \tau_k A_k}{(\alpha-2) \lambda_k} \sum_{j=1}^{K} { \lambda_j \hat{P}_{jk}^{2 / \alpha} \hat{B}_{jk}^{ 2 / \alpha - 1 } } \nonumber \\
%&\overset{(a)}{=} \frac{-2 \tau_k}{\alpha-2} \sum_{j=1}^K { A_j \hat{B}_{jk}^{-1} },
%\end{align}
\begin{equation} \label{Eq:ELogCDownlinkShared4}
\mathbb{E}_{\left|\bx_{k,0}\right|,\Phi,\mathbf{g}} \left[ \log \left( C_k \right) \right] = \frac{-2 \tau_k A_k}{(\alpha-2) \lambda_k} \sum_{j=1}^{K} { \lambda_j \hat{P}_{jk}^{2 / \alpha} \hat{B}_{jk}^{ 2 / \alpha - 1 } } \overset{(a)}{=} \frac{-2 \tau_k}{\alpha-2} \sum_{j=1}^K { A_j \hat{B}_{jk}^{-1} },
\end{equation}
where in (a), we notice from (\ref{Eq:AssociationProbability}) that
\begin{equation} \label{Eq:AssociationProbability2}
\hat{A}_{jk} = \hat{\lambda}_{jk} \left( \hat{P}_{jk} \hat{B}_{jk} \right)^{2/\alpha} \Rightarrow \frac{\lambda_k}{A_k} = \frac{\lambda_j}{A_j} \left( \hat{P}_{jk} \hat{B}_{jk} \right)^{2/\alpha}.
\end{equation}

\subsubsection{Orthogonal Spectrum Partition}

The SIR of the user associated with the $k$-th tier is
\begin{equation}
\text{SIR}_k = \frac{ P_k \left|\bx_{k,0}\right|^{-\alpha} g\left(\bx_{k,0},\bzero\right) }{ I_{\Phi_k} },
\end{equation}
where $I_{\Phi_k}$ denotes the interference from BSs in the $k$-th tier
\begin{equation}
I_{\Phi_k} = \sum_{ i: \bx_{k,i} \in \Phi_k \backslash \bx_{k,0} } { P_k \left|\bx_{k,i}\right|^{-\alpha} g\left(\bx_{k,i},\bzero\right) }.
\end{equation}

Similar to the discussion for the spectrum sharing case, the mean logarithm of the coverage probability for downlink spectrum partition case can be derived, without cross-tier interference involved, as
\begin{equation} \label{Eq:ELogCDownlinkOrthogonal}
\mathbb{E}_{\left|\bx_{k,0}\right|,\Phi_k,\mathbf{g}} \left[ \log \left( C_k \right) \right] = \frac{-2 \tau_k A_k}{\alpha-2}.
\end{equation}

\subsection{Mean Logarithm of Coverage Probability in the Uplink}

We derive the mean logarithm of coverage probability of the typical user in the uplink, assuming that the typical BS is in the $k$-th tier. We re-define the coordinates so that the typical BS is located at $\bx_{k,0} = \bzero$. Both the spectrum sharing and orthogonal spectrum partition schemes are considered.

Under the fractional power control model as described in Section II, the received signal power at the typical BS is
\begin{equation} \label{Eq:UplinkPowerControl}
P_{k,0}^{(r)} = P_u \left|\by_{k,0}\right|^{(\epsilon - 1) \alpha} g\left(\by_{k,0},\bzero\right).
\end{equation}
The sum interference received at the typical BS on a given spectrum resource block comes from users scheduled by other BSs in that spectrum. The interference from user $(j,i)$ is
\begin{equation}
P_{j,i}^{(r)} = P_u \left|\by_{j,i}-\bx_{j,i}\right|^{\epsilon \alpha} \left|\by_{j,i}\right|^{-\alpha} g\left(\by_{j,i},\bzero\right).
\end{equation}
The uplink transmit power $P_u \left|\by_{j,i}-\bx_{j,i}\right|^{\epsilon \alpha}$ of user $(j,i)$ can be modeled as a random variable since the distance $\left|\by_{j,i}-\bx_{j,i}\right|$ is different for each scheduled user. From (\ref{Eq:AssociationRule}), given that user $(j,i)$ is associated with a tier-$j$ BS, we have the following inequality
\begin{equation} \label{Eq:UplinkLengthConstraint}
\left|\by_{j,i}-\bx_{j,i}\right| < \left|\by_{j,i}\right| \left( \hat{P}_{jk} \hat{B}_{jk} \right)^{1/\alpha}, \; \forall (j,i) \neq (k,0).
\end{equation}

\subsubsection{Spectrum Sharing}

The SIR of the user associated with the $k$-th tier is
\begin{equation}
\text{SIR}_k = \frac{ P_u \left|\by_{k,0}\right|^{(\epsilon - 1) \alpha} g\left(\by_{k,0},\bzero\right) }{ I_{\Psi'} },
\end{equation}
where $I_{\Psi'}$ denotes the interference from all the scheduled users
\begin{equation}
I_{\Psi'} = \sum_{j=1}^K \sum_{ i: \by_{j,i} \in \Psi_j \backslash \by_{k,0} } { P_u \left|\by_{j,i}-\bx_{j,i}\right|^{\epsilon \alpha} \left|\by_{j,i}\right|^{-\alpha} g\left(\by_{j,i},\bzero\right) },
\end{equation}
Here, the interference is summed over the set of users $\Psi_j$ scheduled by the $j$-th tier, excluding the typical user.

The mean logarithm of the coverage probability is averaged over the locations of the typical user and interfering users, the location of BSs, and the interference channel:
%\begin{align} \label{Eq:ELogCUplinkShared}
%&\mathbb{E}_{\left|\by_{k,0}\right|,\Psi',\Phi,\mathbf{g}} \left[ \log \left( C_k \right) \right] \nonumber \\
%=& \mathbb{E}_{\left|\by_{k,0}\right|,\Psi',\Phi,\mathbf{g}} \left\{ \log \left[ \mathbb{P} \left( \text{SIR}_k > \tau_k \right) \right] \right\} \nonumber \\
%=& \mathbb{E}_{\left|\by_{k,0}\right|,\Psi',\Phi,\mathbf{g}} \left\{ \log \left[ \mathbb{P} \hspace{-0.5mm} \left( \hspace{-0.5mm} g\left(\by_{k,0},\bzero\right) \hspace{-0.5mm} > \hspace{-0.5mm} \tau_k P_u^{-1} \hspace{-0.5mm} \left|\by_{k,0}\right|^{(1 - \epsilon) \alpha} \hspace{-0.5mm} I_{\Psi'} \hspace{-0.5mm} \right) \right] \right\} \nonumber \\
%=& \mathbb{E}_{\left|\by_{k,0}\right|,\Psi',\Phi,\mathbf{g}} \left\{ \log \left[ \exp \left( - \tau_k P_u^{-1} \left|\by_{k,0}\right|^{(1 - \epsilon) \alpha} I_{\Psi'} \right) \right] \right\} \nonumber \\
%\approx& - \tau_k P_u^{-1} \mathbb{E}_{\left|\by_{k,0}\right|} \left( \left|\by_{k,0}\right|^{(1 - \epsilon) \alpha} \right) \mathbb{E}_{\Psi',\Phi,\mathbf{g}} \left( I_{\Psi'} \right).
%\end{align}
\begin{align} \label{Eq:ELogCUplinkShared}
\mathbb{E}_{\left|\by_{k,0}\right|,\Psi',\Phi,\mathbf{g}} \left[ \log \left( C_k \right) \right] &= \mathbb{E}_{\left|\by_{k,0}\right|,\Psi',\Phi,\mathbf{g}} \left\{ \log \left[ \mathbb{P} \left( \text{SIR}_k > \tau_k \right) \right] \right\} \nonumber \\
&= \mathbb{E}_{\left|\by_{k,0}\right|,\Psi',\Phi,\mathbf{g}} \left\{ \log \left[ \mathbb{P} \hspace{-0.5mm} \left( \hspace{-0.5mm} g\left(\by_{k,0},\bzero\right) \hspace{-0.5mm} > \hspace{-0.5mm} \tau_k P_u^{-1} \hspace{-0.5mm} \left|\by_{k,0}\right|^{(1 - \epsilon) \alpha} \hspace{-0.5mm} I_{\Psi'} \hspace{-0.5mm} \right) \right] \right\} \nonumber \\
&= \mathbb{E}_{\left|\by_{k,0}\right|,\Psi',\Phi,\mathbf{g}} \left\{ \log \left[ \exp \left( - \tau_k P_u^{-1} \left|\by_{k,0}\right|^{(1 - \epsilon) \alpha} I_{\Psi'} \right) \right] \right\} \nonumber \\
&\approx - \tau_k P_u^{-1} \mathbb{E}_{\left|\by_{k,0}\right|} \left( \left|\by_{k,0}\right|^{(1 - \epsilon) \alpha} \right) \mathbb{E}_{\Psi',\Phi,\mathbf{g}} \left( I_{\Psi'} \right).
\end{align}
Note that unlike the downlink case, the length of the uplink interfering links is not lower bounded by a function of the serving link length $\left|\by_{k,0}\right|$, i.e., the interfering users can be anywhere irrespective of the typical user location. However, there is still dependency between $\left|\by_{k,0}\right|$ and the uplink interference $I_{\Psi'}$. For analytical tractability, the derivation here ignores such dependency and approximates $\mathbb{E} \left( \left|\by_{k,0}\right|^{(1 - \epsilon) \alpha} I_{\Psi'} \right)$ by $\mathbb{E} \left( \left|\by_{k,0}\right|^{(1 - \epsilon) \alpha} \right) \mathbb{E} \left( I_{\Psi'} \right)$ in the last step of (\ref{Eq:ELogCUplinkShared}).

Denoting $d_{j,i} = \left|\by_{j,i}-\bx_{j,i}\right|$ for notational simplicity, the uplink mean interference is
%\begin{align} \label{Eq:MeanInterferenceUplinkShared1}
%&\mathbb{E}_{\Psi',\Phi,\mathbf{g}} \left( I_{\Psi'} \right) \nonumber \\
%=& P_u \sum_{j=1}^K \mathbb{E}_{\Psi_j,\Phi_j} \left( \sum_{ i: \by_{j,i} \in \Psi_j \backslash \by_{k,0} } d_{j,i}^{\epsilon \alpha} \left|\by_{j,i}\right|^{-\alpha} \right) \nonumber \\
%=& P_u \sum_{j=1}^K \mathbb{E}_{\Psi_j,\Phi_j} \left[ \sum_{ i: \by_{j,i} \in \Psi_j } d_{j,i}^{\epsilon \alpha} \left|\by_{j,i}\right|^{-\alpha} \mathbf{1}\left( \by_{j,i} \neq \by_{k,0} \right) \right] \nonumber \\
%\overset{(a)}{=}& P_u \hspace{-0.5mm} \sum_{j=1}^K 2\pi \lambda_j \hspace{-0.5mm} \int_0^\infty \hspace{-2mm} \mathbb{E}_{d_{j,i}} \hspace{-1mm} \left[ d_{j,i}^{\epsilon \alpha} \mathbf{1} \hspace{-0.5mm} \left( \by_{j,i} \hspace{-0.5mm} \neq \hspace{-0.5mm} \by_{k,0} \right) \hspace{-0.5mm} \big| \left|\by_{j,i}\right| \hspace{-0.5mm} = \hspace{-0.5mm} y \right] y^{-\alpha} y \text{d}y \nonumber \\
%\overset{(b)}{=}& 2\pi P_u \hspace{-0.5mm} \sum_{j=1}^K \hspace{-0.5mm} \lambda_j \hspace{-0.5mm} \int_0^\infty \hspace{-2mm} \mathbb{E}_{d_{j,i}} \hspace{-1mm} \left\{ \hspace{-0.5mm} d_{j,i}^{\epsilon \alpha} \mathbf{1} \hspace{-1mm} \left[ d_{j,i} \hspace{-0.5mm} < y \hspace{-0.5mm} \left( \hat{P}_{jk} \hat{B}_{jk} \right)^{\hspace{-0.5mm}1/\alpha} \right] \hspace{-0.5mm} \right\} \hspace{-0.5mm} y^{1-\alpha} \text{d}y.
%\end{align}
\begin{align} \label{Eq:MeanInterferenceUplinkShared1}
\mathbb{E}_{\Psi',\Phi,\mathbf{g}} \left( I_{\Psi'} \right) &= P_u \sum_{j=1}^K \mathbb{E}_{\Psi_j,\Phi_j} \left( \sum_{ i: \by_{j,i} \in \Psi_j \backslash \by_{k,0} } d_{j,i}^{\epsilon \alpha} \left|\by_{j,i}\right|^{-\alpha} \right) \nonumber \\
&= P_u \sum_{j=1}^K \mathbb{E}_{\Psi_j,\Phi_j} \left[ \sum_{ i: \by_{j,i} \in \Psi_j } d_{j,i}^{\epsilon \alpha} \left|\by_{j,i}\right|^{-\alpha} \mathbf{1}\left( \by_{j,i} \neq \by_{k,0} \right) \right] \nonumber \\
&\overset{(a)}{=} P_u \hspace{-0.5mm} \sum_{j=1}^K 2\pi \lambda_j \hspace{-0.5mm} \int_0^\infty \hspace{-2mm} \mathbb{E}_{d_{j,i}} \hspace{-1mm} \left[ d_{j,i}^{\epsilon \alpha} \mathbf{1} \hspace{-0.5mm} \left( \by_{j,i} \hspace{-0.5mm} \neq \hspace{-0.5mm} \by_{k,0} \right) \hspace{-0.5mm} \big| \left|\by_{j,i}\right| \hspace{-0.5mm} = \hspace{-0.5mm} y \right] y^{-\alpha} y \text{d}y \nonumber \\
&\overset{(b)}{=} 2\pi P_u \hspace{-0.5mm} \sum_{j=1}^K \hspace{-0.5mm} \lambda_j \hspace{-0.5mm} \int_0^\infty \hspace{-2mm} \mathbb{E}_{d_{j,i}} \hspace{-1mm} \left\{ \hspace{-0.5mm} d_{j,i}^{\epsilon \alpha} \mathbf{1} \hspace{-1mm} \left[ d_{j,i} \hspace{-0.5mm} < y \hspace{-0.5mm} \left( \hat{P}_{jk} \hat{B}_{jk} \right)^{\hspace{-0.5mm}1/\alpha} \right] \hspace{-0.5mm} \right\} \hspace{-0.5mm} y^{1-\alpha} \text{d}y.
\end{align}
In (a) we use Campbell's Formula\footnote{Campbell's Formula does not need the scheduled user set $\Psi_j$ to be a PPP; it can be applied as long as $\Psi_j$ is a point process with a finite intensity.}. The intensity of $\Psi_j$ is the same as that of the BS PPP $\Phi_j$, as only one user is scheduled by each BS at a time on each resource block\footnote{In some realizations there may be no users in a cell, and hence no users are scheduled. If $\lambda_u \gg \lambda_j$, such event hardly occurs.}. In (b) we use the inequality (\ref{Eq:UplinkLengthConstraint}) for the interfering users. Unlike the downlink, the integral starts from $0$ as interfering users can be arbitrarily close to the typical BS, as long as their associated BSs could potentially be located arbitrarily close to the typical BS.

The distance $d_{j,i}$ follows the same distribution as $\left|\bx_{k,0}\right|$ in (\ref{Eq:DistanceDistribution}) with the index changed from $k$ to $j$
\begin{equation} \label{Eq:ConditionalDistanceDistribution}
f_{d_{j,i}}(r) =  2 \pi r \frac{\lambda_j}{A_j} \exp \left( -\pi r^2 \frac{\lambda_j}{A_j} \right), \quad r \ge 0.
\end{equation}
Using (\ref{Eq:ConditionalDistanceDistribution}) in (\ref{Eq:MeanInterferenceUplinkShared1}), we have
%\begin{align} \label{Eq:MeanInterferenceUplinkShared2}
%&\mathbb{E}_{\Psi',\Phi,\mathbf{g}} \left( I_{\Psi'} \right) = 4 \pi^2 P_u \sum_{j=1}^K \vast[ \frac{\lambda_j^2}{A_j} \times \nonumber \\
%&\int_0^\infty y^{1-\alpha} \int_0^{y \left( \hat{P}_{jk} \hat{B}_{jk} \right)^{1/\alpha}} r^{\epsilon \alpha + 1} \exp \left( -\pi r^2 \frac{\lambda_j}{A_j} \right) \text{d}r \text{d}y \vast] \nonumber \\
%\overset{(a)}{=}& 4 \pi P_u \sum_{j=1}^K \vast[ \lambda_j \left( \frac{A_j}{\pi \lambda_j} \right)^{\epsilon \alpha / 2} \times \nonumber \\
%&\int_0^\infty y^{1-\alpha} \int_0^{y \sqrt{\pi \lambda_j / A_j} \left( \hat{P}_{jk} \hat{B}_{jk} \right)^{1/\alpha}} u^{\epsilon \alpha + 1} e^{-u^2} \text{d}u \text{d}y \vast] \nonumber \\
%\overset{(b)}{=}& 4 \pi P_u \left( \frac{A_k}{\pi \lambda_k} \right)^{\epsilon \alpha / 2} \sum_{j=1}^K \lambda_j \left( \hat{P}_{jk} \hat{B}_{jk} \right)^\epsilon \times \nonumber \\
%&\int_0^\infty y^{1-\alpha} \int_0^{y \sqrt{\pi \lambda_k / A_k}} u^{\epsilon \alpha + 1} e^{-u^2} \text{d}u \text{d}y \nonumber \\
%\overset{(c)}{=}& 4 \pi P_u \Xi (\alpha, \epsilon) \left( \frac{A_k}{\pi \lambda_k} \right)^{\hspace{-1mm} 1 + (\epsilon-1) \alpha / 2} \sum_{j=1}^K \lambda_j \left( \hat{P}_{jk} \hat{B}_{jk} \right)^\epsilon,
%\end{align}
\begin{align} \label{Eq:MeanInterferenceUplinkShared2}
\mathbb{E}_{\Psi',\Phi,\mathbf{g}} \left( I_{\Psi'} \right) &= 4 \pi^2 P_u \sum_{j=1}^K \vast[ \frac{\lambda_j^2}{A_j} \int_0^\infty y^{1-\alpha} \int_0^{y \left( \hat{P}_{jk} \hat{B}_{jk} \right)^{1/\alpha}} r^{\epsilon \alpha + 1} \exp \left( -\pi r^2 \frac{\lambda_j}{A_j} \right) \text{d}r \text{d}y \vast] \nonumber \\
&\overset{(a)}{=} 4 \pi P_u \sum_{j=1}^K \vast[ \lambda_j \left( \frac{A_j}{\pi \lambda_j} \right)^{\epsilon \alpha / 2} \int_0^\infty y^{1-\alpha} \int_0^{y \sqrt{\pi \lambda_j / A_j} \left( \hat{P}_{jk} \hat{B}_{jk} \right)^{1/\alpha}} u^{\epsilon \alpha + 1} e^{-u^2} \text{d}u \text{d}y \vast] \nonumber \\
&\overset{(b)}{=} 4 \pi P_u \left( \frac{A_k}{\pi \lambda_k} \right)^{\epsilon \alpha / 2} \sum_{j=1}^K \lambda_j \left( \hat{P}_{jk} \hat{B}_{jk} \right)^\epsilon \int_0^\infty y^{1-\alpha} \int_0^{y \sqrt{\pi \lambda_k / A_k}} u^{\epsilon \alpha + 1} e^{-u^2} \text{d}u \text{d}y \nonumber \\
&\overset{(c)}{=} 4 \pi P_u \Xi (\alpha, \epsilon) \left( \frac{A_k}{\pi \lambda_k} \right)^{\hspace{-1mm} 1 + (\epsilon-1) \alpha / 2} \sum_{j=1}^K \lambda_j \left( \hat{P}_{jk} \hat{B}_{jk} \right)^\epsilon,
\end{align}
where (a) follows from the change of variables $u = r \sqrt{\frac{\pi \lambda_j}{A_j}}$, in (b) we use (\ref{Eq:AssociationProbability2}), (c) follows from the change of variables $v = y \sqrt{\frac{\pi \lambda_k}{A_k}}$, and the function $\Xi (\alpha, \epsilon)$ in (\ref{Eq:MeanInterferenceUplinkShared2}) is defined as
\begin{align} \label{Eq:Xi}
\Xi (\alpha, \epsilon) &\triangleq \int_0^\infty v^{1-\alpha} \int_0^v u^{\epsilon \alpha + 1} e^{-u^2} \text{d}u \text{d}v \nonumber \\
&= \int_0^\infty u^{\epsilon \alpha + 1} e^{-u^2} \int_u^\infty v^{1-\alpha} \text{d}v \text{d}u \nonumber \\
&= \frac{1}{\alpha-2} \int_0^\infty u^{3+(\epsilon-1)\alpha} e^{-u^2} \text{d}u \nonumber \\
&= \frac{1}{2(\alpha-2)} \Gamma \left[ 2 + \frac{(\epsilon-1)\alpha}{2} \right].
\end{align}
Note that for the integral to converge, one requires the parameter of the Gamma function $\Gamma(\cdot)$ to be positive, i.e., the power control factor $\epsilon$ needs to satisfy the following constraints
\begin{equation} \label{Eq:PowerControlConstraint}
\epsilon > 1 - \frac{4}{\alpha} \quad \text{and} \quad 0 \le \epsilon \le 1.
\end{equation}
The physical interpretation behind this is that, the uplink mean interference is unbounded when $\epsilon$ is not greater than $1 - \frac{4}{\alpha}$, as interfering users can be arbitrarily close to the typical BS.

Now, recognizing again that the PDF of the distance $\left|\by_{k,0}\right|$ between the typical BS and the typical user is the same as (\ref{Eq:DistanceDistribution}) (with coordinates shifted), the first expectation in (\ref{Eq:ELogCUplinkShared}) can be expressed as
%\begin{align} \label{Eq:MeanSignalPowerInverseUplink}
%\mathbb{E}_{\left|\by_{k,0}\right|} \hspace{-0.5mm} \left( \left|\by_{k,0}\right|^{(1 - \epsilon) \alpha} \hspace{-0.5mm} \right) \hspace{-0.5mm} &= 2\pi \frac{\lambda_k}{A_k} \hspace{-1mm} \int_0^\infty \hspace{-2mm} r^{1 + (1-\epsilon)\alpha} \hspace{-0.5mm} \exp \hspace{-1mm} \left( \hspace{-0.5mm} -\pi r^2 \frac{\lambda_k}{A_k} \hspace{-0.5mm} \right) \hspace{-0.5mm} \text{d}r \nonumber \\
%&\overset{(a)}{=} 2 \Upsilon (\alpha, \epsilon) \left( \frac{A_k}{\pi \lambda_k} \right)^{(1-\epsilon) \alpha / 2},
%\end{align}
\begin{equation} \label{Eq:MeanSignalPowerInverseUplink}
\mathbb{E}_{\left|\by_{k,0}\right|} \hspace{-0.5mm} \left( \left|\by_{k,0}\right|^{(1 - \epsilon) \alpha} \hspace{-0.5mm} \right) \hspace{-0.5mm} = 2\pi \frac{\lambda_k}{A_k} \hspace{-1mm} \int_0^\infty \hspace{-2mm} r^{1 + (1-\epsilon)\alpha} \hspace{-0.5mm} \exp \hspace{-1mm} \left( \hspace{-0.5mm} -\pi r^2 \frac{\lambda_k}{A_k} \hspace{-0.5mm} \right) \hspace{-0.5mm} \text{d}r \overset{(a)}{=} 2 \Upsilon (\alpha, \epsilon) \left( \frac{A_k}{\pi \lambda_k} \right)^{(1-\epsilon) \alpha / 2},
\end{equation}
where in (a) we use the change of variables $t = r \sqrt{\frac{\pi \lambda_k}{A_k}}$ and define the function $\Upsilon (\alpha, \epsilon)$ as
\begin{equation} \label{Eq:Upsilon}
\Upsilon (\alpha, \epsilon) \triangleq \int_0^\infty t^{1 + (1-\epsilon)\alpha} e^{-t^2} \text{d}t = \frac{1}{2} \Gamma \left[ 1 + \frac{(1-\epsilon)\alpha}{2} \right].
\end{equation}

Combining (\ref{Eq:MeanSignalPowerInverseUplink}) and (\ref{Eq:MeanInterferenceUplinkShared2}) and substituting into (\ref{Eq:ELogCUplinkShared}),
%\begin{align} \label{Eq:ELogCUplinkShared2}
%&\mathbb{E}_{\left|\by_{k,0}\right|,\Psi',\Phi,\mathbf{g}} \left[ \log \left( C_k \right) \right] \nonumber \\
%\approx& -8 \tau_k \Upsilon (\alpha, \epsilon) \Xi (\alpha, \epsilon) \frac{A_k}{\lambda_k} \sum_{j=1}^K \lambda_j \left( \hat{P}_{jk} \hat{B}_{jk} \right)^\epsilon \nonumber \\
%=& -8 \tau_k \Omega (\alpha, \epsilon) \sum_{j=1}^K A_j \left( \hat{P}_{jk} \hat{B}_{jk} \right)^{\epsilon - 2/\alpha},
%\end{align}
\begin{align} \label{Eq:ELogCUplinkShared2}
\mathbb{E}_{\left|\by_{k,0}\right|,\Psi',\Phi,\mathbf{g}} \left[ \log \left( C_k \right) \right] &\approx -8 \tau_k \Upsilon (\alpha, \epsilon) \Xi (\alpha, \epsilon) \frac{A_k}{\lambda_k} \sum_{j=1}^K \lambda_j \left( \hat{P}_{jk} \hat{B}_{jk} \right)^\epsilon \nonumber \\
&= -8 \tau_k \Omega (\alpha, \epsilon) \sum_{j=1}^K A_j \left( \hat{P}_{jk} \hat{B}_{jk} \right)^{\epsilon - 2/\alpha},
\end{align}
where we use (\ref{Eq:AssociationProbability2}) in (a) and define $\Omega (\alpha, \epsilon) \triangleq \Upsilon (\alpha, \epsilon) \Xi (\alpha, \epsilon)$.

\subsubsection{Orthogonal Spectrum Partition}

The SIR of the user associated with the $k$-th tier is
\begin{equation}
\text{SIR}_k = \frac{ P_u \left|\by_{k,0}\right|^{(\epsilon - 1) \alpha} g\left(\by_{k,0},\bzero\right) }{ I_{\Psi_k} },
\end{equation}
where $I_{\Psi_k}$ denotes the interference from the scheduled users in the $k$-th tier
\begin{equation}
I_{\Psi_k} = \sum_{ i: \by_{k,i} \in \Psi_k \backslash \by_{k,0} } { P_u \left|\by_{k,i}-\bx_{k,i}\right|^{\epsilon \alpha} \left|\by_{k,i}\right|^{-\alpha} g\left(\by_{k,i},\bzero\right) }.
\end{equation}

The mean logarithm of the coverage probability can be derived similarly, without cross-tier interference, as
\begin{equation} \label{Eq:ELogCUplinkOrthogonal}
\mathbb{E}_{\left|\by_{k,0}\right|,\Psi_k,\Phi_k,\mathbf{g}} \left[ \log \left( C_k \right) \right] = -8 \tau_k \Omega (\alpha, \epsilon) A_k.
\end{equation}

\section{Utility Optimization}

This section presents the optimization of the derived mean user utility. We consider both spectrum sharing and orthogonal spectrum partition schemes for both downlink and uplink.

\subsection{Spectrum Sharing in the Downlink}

Combining (\ref{Eq:MeanPerUserUtility}), (\ref{Eq:UtilityPerTier}), (\ref{Eq:PerUserResource}), and (\ref{Eq:ELogCDownlinkShared4}), considering that $W_k = W$ under spectrum sharing, the downlink mean utility is
\begin{equation} \label{Eq:MeanPerUserUtilityDownlinkShared}
U = \sum_{k=1}^K \hspace{-0.5mm} A_k \hspace{-0.5mm} \left\{ \hspace{-0.5mm} \log \left[ \frac{W \lambda_k \log (1 + \tau_k)}{A_k \lambda_u} \right] \hspace{-0.5mm} - \hspace{-0.5mm} \frac{2 \tau_k}{\alpha - 2} \sum_{j=1}^K { A_j \hat{B}_{jk}^{-1} } \hspace{-0.5mm} \right\}.
\end{equation}

We only need to find the optimal bias factors to maximize the mean user utility. Instead of directly optimizing over $\left\{ B_k \right\}_{\forall k}$, we optimize over the association probability $\left\{ A_k \right\}_{\forall k}$: from (\ref{Eq:AssociationProbability2}) we substitute $\hat{B}_{jk} = \hat{P}_{jk}^{-1} \hat{\lambda}_{jk}^{-\alpha/2} \hat{A}_{jk}^{\alpha/2}$ into (\ref{Eq:MeanPerUserUtilityDownlinkShared}) and formulate the mean utility as
%\begin{align} \label{Eq:MeanPerUserUtilityDownlinkShared3}
%&U \left( \left\{A_k\right\}_{\forall k} \right) = \sum_{k=1}^K A_k \log \left[ \frac{W \lambda_k \log (1 + \tau_k)}{A_k \lambda_u} \right] \nonumber \\
%&\quad\quad - \frac{2}{\alpha - 2} \sum_{i=1}^K \frac{\tau_i}{P_i \lambda_i^{\alpha/2}} A_i^{1+\alpha/2} \sum_{j=1}^K P_j \lambda_j^{\alpha/2} A_j^{1-\alpha/2},
%\end{align}
\begin{equation} \label{Eq:MeanPerUserUtilityDownlinkShared3}
U \left( \left\{A_k\right\}_{\forall k} \right) = \sum_{k=1}^K A_k \log \left[ \frac{W \lambda_k \log (1 + \tau_k)}{A_k \lambda_u} \right] - \frac{2}{\alpha - 2} \sum_{i=1}^K \frac{\tau_i}{P_i \lambda_i^{\alpha/2}} A_i^{1+\alpha/2} \sum_{j=1}^K P_j \lambda_j^{\alpha/2} A_j^{1-\alpha/2},
\end{equation}
and the utility maximization problem becomes
\begin{subequations} \label{Eq:OptimalAssociationProblemDownlinkShared}
\begin{eqnarray}
&\displaystyle \mathop\text{maximize}\limits_{A_k,\forall k} & \; U \left( \left\{A_k\right\}_{\forall k} \right), \\
&\text{subject to} & \; \sum_{k=1}^K { A_k } = 1, \label{Eq:OptimalAssociationProblemDownlinkShared2}\\
&& A_k > 0, \; \forall k. \label{Eq:OptimalAssociationProblemDownlinkShared3}
\end{eqnarray}
\end{subequations}
This problem does not have a closed-form solution, and it is not convex in general. However, numerical solutions can be obtained efficiently to arrive at a local optimum. The effectiveness of such numerical approach is validated in simulations. Finally, using (\ref{Eq:BasA}) we can recover the bias factors of each tier $\left\{ B_k^* \right\}_{\forall k}$ from $\left\{ A_k^* \right\}_{\forall k}$.

\subsection{Orthogonal Spectrum Partition in the Downlink}

With $W_k = W \eta_k$ under spectrum partition, and substituting (\ref{Eq:ELogCDownlinkOrthogonal}) in (\ref{Eq:UtilityPerTier}), the downlink mean utility is
\begin{equation} \label{Eq:MeanPerUserUtilityDownlinkOrthogonal}
U = \sum_{k=1}^K A_k \left\{ \log \left[ \frac{W \eta_k \lambda_k \log (1 + \tau_k)}{A_k \lambda_u} \right] - \frac{2 \tau_k A_k}{\alpha - 2} \right\}.
\end{equation}

\subsubsection{Optimal Spectrum Partition}

The optimal spectrum partition problem is formulated as
\begin{subequations} \label{Eq:OptimalSpectrumPartitionProblem}
\begin{eqnarray}
&\displaystyle \mathop\text{maximize}\limits_{\eta_k, \forall k} & \; U \left( \left\{ \eta_k \right\}_{\forall k} \right), \\
&\text{subject to} & \; \sum_{k=1}^K { \eta_k } = 1, \label{Eq:OptimalSpectrumPartitionProblem1}\\
&& \eta_k > 0, \; \forall k.
\end{eqnarray}
\end{subequations}

Solving problem (\ref{Eq:OptimalSpectrumPartitionProblem}) leads to the following theorem.
\begin{thm}
For downlink under orthogonal spectrum partition, the optimal proportion of spectrum allocated to a tier is equal to the proportion of users associated with that tier.
\end{thm}
\begin{IEEEproof}
By introducing the dual variable $\mu$ with respect to the constraint (\ref{Eq:OptimalSpectrumPartitionProblem1}), we form the Lagrangian
\begin{equation}
g \left( \mu \right) = U \left( \left\{ \eta_k \right\}_{\forall k} \right) - \mu \left( \sum_{k=1}^K { \eta_k } - 1 \right).
\end{equation}
The Karush-Kuhn-Tucker (KKT) condition can be obtained via taking the first order derivative with respect to $\eta_k$ as
\begin{equation}
\frac{A_k}{\eta_k^*} - \mu = 0.
\end{equation}
Since $\sum_{k=1}^K { \eta_k^* } = \frac{\sum_{k=1}^K { A_k }}{\mu} = \frac{1}{\mu} = 1$, we have $\mu = 1$ and consequently the optimal spectrum partition follows
\begin{equation} \label{Eq:OptimalSpectrumPartitionDownlink}
\eta_k^* = A_k.
\end{equation}
Since utility (\ref{Eq:MeanPerUserUtilityDownlinkOrthogonal}) is concave in $\eta_k$, $\eta_k^*$ achieves maximum. Note that the user association probability of a tier is equivalent to the mean proportion of users associated with that tier.
\end{IEEEproof}

\subsubsection{Optimal User Association}

Let $\eta_k^* = A_k$ in (\ref{Eq:MeanPerUserUtilityDownlinkOrthogonal}) and reformulate the problem (\ref{Eq:OptimalAssociationProblemDownlinkShared}) to maximize the utility
\begin{equation} \label{Eq:OptimalAssociationProblemDownlinkOrthogonal}
U \left( \left\{ A_k \right\}_{\forall k} \right) \hspace{-0.5mm} = \hspace{-0.5mm} \sum_{k=1}^K \hspace{-0.5mm} A_k \hspace{-0.5mm} \left\{ \hspace{-0.5mm} \log \hspace{-0.5mm} \left[ \hspace{-0.5mm} \frac{W \lambda_k \log\left( 1 + \tau_k \right)}{\lambda_u} \hspace{-0.5mm} \right] - \frac{2 \tau_k A_k}{\alpha-2} \hspace{-0.5mm} \right\},
\end{equation}
subject to the constraints in (\ref{Eq:OptimalAssociationProblemDownlinkShared}). We get Theorem 2 as follows.
\begin{thm}
For downlink under orthogonal spectrum partition, the optimal user association bias $\left\{ B_k^* \right\}_{\forall k}$ can be obtained via (\ref{Eq:BasA}) from the optimal $\left\{ A_k^* \right\}_{\forall k}$, which is given as
\begin{equation} \label{Eq:OptimalAssociationDownlinkOrthogonal}
A_k^* = \max \left\{ \log \left[ \frac{W \lambda_k \log\left( 1 + \tau_k \right)}{\lambda_u} \right] - \nu, 0 \right\} \frac{\alpha - 2}{4 \tau_k},
\end{equation}
where $\nu$ is chosen such that $\sum_{k=1}^K { A_k^* } = 1$ is satisfied.
\end{thm}
\begin{IEEEproof}
Employing the Lagrangian method to maximize (\ref{Eq:OptimalAssociationProblemDownlinkOrthogonal}) we have
\begin{equation}
q \left( \nu \right) = U \left( \left\{ A_k \right\}_{\forall k} \right) - \nu \left( \sum_{k=1}^K { A_k } - 1 \right),
\end{equation}
where $\nu$ is the corresponding dual variable. The first order condition with respect to $A_k$ is
\begin{equation}
\log \left[ \frac{W \lambda_k \log\left( 1 + \tau_k \right)}{\lambda_u} \right] - \frac{4 \tau_k A_k}{\alpha - 2} - \nu = 0,
\end{equation}
and simple manipulations lead to (\ref{Eq:OptimalAssociationDownlinkOrthogonal}). $A_k^*$ achieves global optimum, since the objective (\ref{Eq:OptimalAssociationProblemDownlinkOrthogonal}) is concave in $A_k$, $\forall k$.
\end{IEEEproof}
The solution in (\ref{Eq:OptimalAssociationDownlinkOrthogonal}) can also be written as $A_k^* = \max \left\{ \log \left[ \frac{\lambda_k \log\left( 1 + \tau_k \right)}{\theta} \right], 0 \right\} \frac{\alpha - 2}{4 \tau_k}$. This means that whenever the value of $\lambda_k \log\left( 1 + \tau_k \right)$ of tier-$k$ is above some threshold $\theta$, the optimal association probability to this tier is proportional to $\log \lambda_k$ and roughly inversely proportional to $\tau_k$. Otherwise, no users should associate with tier-$k$ as far as maximizing the proportionally fair utility is concerned. Intuitively, users tend to associate with BS tiers with larger deployment intensity as the access distance is shorter. Users favouring tiers with lower target SIR implies that the raised coverage probability by decreasing target SIR offsets the correspondingly reduced rate and benefits utility.

Finally, the optimal spectrum partition is $\eta_k^* = A_k^*, \forall k$.

\subsection{Spectrum Sharing in the Uplink}

Substituting (\ref{Eq:ELogCUplinkShared2}) in (\ref{Eq:UtilityPerTier}) and considering $W_k = W$ for spectrum sharing, the uplink mean utility is
%\begin{align} \label{Eq:MeanPerUserUtilityUplinkShared}
%U = \sum_{k=1}^K A_k & \vast\{ \log \left[ \frac{W \lambda_k \log (1 + \tau_k)}{A_k \lambda_u} \right] \nonumber \\
%&- 8 \tau_k \Omega (\alpha, \epsilon) \sum_{j=1}^K A_j \left( \hat{P}_{jk} \hat{B}_{jk} \right)^{\epsilon - 2/\alpha} \vast\}.
%\end{align}
\begin{equation} \label{Eq:MeanPerUserUtilityUplinkShared}
U = \sum_{k=1}^K A_k \vast\{ \log \left[ \frac{W \lambda_k \log (1 + \tau_k)}{A_k \lambda_u} \right] - 8 \tau_k \Omega (\alpha, \epsilon) \sum_{j=1}^K A_j \left( \hat{P}_{jk} \hat{B}_{jk} \right)^{\epsilon - 2/\alpha} \vast\}.
\end{equation}

From (\ref{Eq:AssociationProbability2}), we substitute $\hat{P}_{jk} \hat{B}_{jk} = \hat{\lambda}_{jk}^{-\alpha/2} \hat{A}_{jk}^{\alpha/2}$ into (\ref{Eq:MeanPerUserUtilityUplinkShared}), and formulate the optimization problem as to maximize the following utility
%\begin{align} \label{Eq:OptimalAssociationProblemUplinkShared}
%&U \left( \left\{A_k\right\}_{\forall k} \right) = \sum_{k=1}^K A_k \log \left[ \frac{W \lambda_k \log (1 + \tau_k)}{A_k \lambda_u} \right] \nonumber \\
%& \quad\;\; - 8 \Omega (\alpha, \epsilon) \sum_{i=1}^K \tau_i \lambda_i^{\epsilon \alpha/2 - 1} A_i^{2 - \epsilon \alpha/2} \sum_{j=1}^K \frac{A_j^{\epsilon \alpha/2}}{\lambda_j^{\epsilon \alpha/2 - 1}},
%\end{align}
\begin{equation} \label{Eq:OptimalAssociationProblemUplinkShared}
U \left( \left\{A_k\right\}_{\forall k} \right) = \sum_{k=1}^K A_k \log \left[ \frac{W \lambda_k \log (1 + \tau_k)}{A_k \lambda_u} \right] - 8 \Omega (\alpha, \epsilon) \sum_{i=1}^K \tau_i \lambda_i^{\epsilon \alpha/2 - 1} A_i^{2 - \epsilon \alpha/2} \sum_{j=1}^K \frac{A_j^{\epsilon \alpha/2}}{\lambda_j^{\epsilon \alpha/2 - 1}},
\end{equation}
subject to the constraints (\ref{Eq:OptimalAssociationProblemDownlinkShared2}) and (\ref{Eq:OptimalAssociationProblemDownlinkShared3}). Again, the problem does not have a closed-form solution and is not convex in general. Local optimum can be numerically computed efficiently. The corresponding uplink bias $\left\{ B_k^* \right\}_{\forall k}$ can then be obtained via (\ref{Eq:BasA}) from the resulting $\left\{ A_k^* \right\}_{\forall k}$.

Under some special cases, this optimization problem has a closed-form solution. We have the following theorem.
\begin{thm}
For uplink under spectrum sharing, if all tiers have the same target SIR, i.e., $\tau_k = \tau$, $\forall k$, and $\epsilon \alpha = 2$ or $\epsilon \alpha \ge 4$, the optimal user association is distance based, i.e., each user communicates with its closest BS.
\end{thm}
\begin{IEEEproof}
The proof can be found in Appendix B.
\end{IEEEproof}
The distance-based association in the uplink can be explained with the following intuition. With power control, a user far from its serving BS transmits at high power, causing large interference to other BSs, especially to nearby small cell BSs. In order to avoid this, a proper uplink association scheme should connect each user to its closest BS, irrespective of parameters such as BS deployment density and power, so that few users are located too far from their serving BSs.

Theorem 3 is theoretically proved only in the regimes $\epsilon \alpha = 2$ and $\epsilon \alpha \ge 4$. It is shown in Appendix B that the objective in (\ref{Eq:OptimalAssociationProblemUplinkShared}) is neither convex nor concave in the regimes $\epsilon \alpha < 2$ and $2 < \epsilon \alpha < 4$. However, in our simulation, we note that the optimal user association is distance based for all feasible regimes of $\epsilon \alpha$ if $\tau_k = \tau$, $\forall k$.

Note that the optimal bias $\left\{ B_k^* \right\}_{\forall k}$ computed for uplink may be different from that for downlink, which implies that users may associate with different BSs for downlink and uplink transmissions. This asymmetry matches the future development trend in HetNets \cite{JeffreyAndrews/2013/Seven,DavidAstely/2013/LTE,HishamElshaer/2014/Downlink}, where the downlink-uplink decoupling is advocated.

\subsection{Orthogonal Spectrum Partition in the Uplink}

With $W_k = W \eta_k$ under spectrum partition and substituting (\ref{Eq:ELogCUplinkOrthogonal}) into (\ref{Eq:UtilityPerTier}), we have the uplink mean utility as
\begin{equation} \label{Eq:MeanPerUserUtilityUplinkOrthogonal}
U = \sum_{k=1}^K A_k \left\{ \log \left[ \frac{W \eta_k \lambda_k \log (1 + \tau_k)}{A_k \lambda_u} \right] - 8 \tau_k \Omega (\alpha, \epsilon) A_k \right\}.
\end{equation}

\subsubsection{Optimal Spectrum Partition}

By maximizing (\ref{Eq:MeanPerUserUtilityUplinkOrthogonal}) under the constraints in (\ref{Eq:OptimalSpectrumPartitionProblem}), the optimal spectrum partition ratio is also found to be $\eta_k^* = A_k$, leading to the following theorem.
\begin{thm}
For uplink under orthogonal spectrum partition, the optimal proportion of spectrum allocated to a tier is equal to the proportion of users associated with that tier.
\end{thm}
\begin{IEEEproof}
The proof is omitted as it is similar to the downlink case stated in Theorem 1.
\end{IEEEproof}

\subsubsection{Optimal User Association}

We substitute $\eta_k^* = A_k$ into (\ref{Eq:MeanPerUserUtilityUplinkOrthogonal}) and reformulate the problem (\ref{Eq:OptimalAssociationProblemDownlinkShared}) to maximize the utility
\begin{equation} \label{Eq:OptimalAssociationProblemUplinkOrthogonal}
U = \sum_{k=1}^K A_k \left\{ \log \left[ \frac{W \lambda_k \log (1 + \tau_k)}{\lambda_u} \right] - 8 \tau_k \Omega (\alpha, \epsilon) A_k \right\},
\end{equation}
subject to the constraints in (\ref{Eq:OptimalAssociationProblemDownlinkShared}). We have the following result.
\begin{thm}
For uplink under orthogonal spectrum partition, the optimal user association bias $\left\{ B_k^* \right\}_{\forall k}$ can be obtained via (\ref{Eq:BasA}) from the optimal $\left\{ A_k^* \right\}_{\forall k}$, which is given as
\begin{equation} \label{Eq:OptimalAssociationUplinkOrthogonal}
A_k^* = \max \left\{ \log \left[ \frac{W \lambda_k \log (1 + \tau_k)}{\lambda_u} \right] - \nu, 0 \right\} \frac{1}{16 \tau_k \Omega (\alpha, \epsilon)},
\end{equation}
where $\nu$ is chosen such that $\sum_{k=1}^K { A_k^* } = 1$ is satisfied.
\end{thm}
\begin{IEEEproof}
The Lagrangian method similar to the downlink case in Theorem 2 is employed, and $A_k^*$ achieves global optimum since the objective (\ref{Eq:OptimalAssociationProblemUplinkOrthogonal}) is concave in $A_k$, $\forall k$.
\end{IEEEproof}
Finally, the optimal spectrum partition is $\eta_k^* = A_k^*, \forall k$.

We also note the following symmetry between uplink system with full power control and downlink system.
\begin{coro}
Under orthogonal spectrum partition, the uplink system with full power control shares the same optimal spectrum partition ratio and optimal user association bias as that of the downlink system given the same network parameters (density, power, target SIR and path loss exponent).
\end{coro}
\begin{IEEEproof}
With $\epsilon=1$, $\Xi(\alpha,\epsilon) = \frac{1}{2(\alpha-2)}$ and $\Upsilon(\alpha,\epsilon) = \frac{1}{2}$. Replacing $\Omega(\alpha,\epsilon) = \Xi(\alpha,\epsilon) \Upsilon(\alpha,\epsilon) = \frac{1}{4(\alpha-2)}$ into the solution in (\ref{Eq:OptimalAssociationUplinkOrthogonal}), we find the resulting $A_k^*$ is the same as that in (\ref{Eq:OptimalAssociationDownlinkOrthogonal}) for downlink. Hence the corresponding $\eta_k^*$ and $B_k^*$ of downlink and uplink are also the same.
\end{IEEEproof}

\section{Numerical Results}

In this section, we present numerical results to demonstrate the effectiveness of the proposed joint optimization of user association bias and spectrum partition in multi-tier HetNets. We set the path loss exponent $\alpha=4$. The user PPP $\Phi_u$ has an intensity $\lambda_u = \frac{100}{\pi (1000m)^2}$. The locations of BSs from different tiers are drawn from PPPs with their given density. The system bandwidth $W = 20$MHz is divided into $2048$ subcarriers. We perform Monte Carlo simulation over 50,000 snapshots of different spatial topologies. Each snapshot consists of 20 time slots. The channel fading coefficients are generated according to i.i.d. Rayleigh distribution over both the time slots and the frequency subcarriers. Round robin user scheduling is adopted. Users can be scheduled on multiple subcarriers, thus the user coverage rate is proportional to the number of subcarriers with SIR larger than the threshold.

We study a system with $K=2$ tiers. Tier-1 consists of macro-BSs with lower deployment intensity and higher transmission power, while tier-2 consists of femto-BSs with higher intensity and lower power. The intensities of the BS PPPs are $\lambda_k = a_k \lambda_u$, where $\left\{ a_1, a_2 \right\} = \left\{ 0.01, 0.09 \right\}$. The transmission power of the two tiers are $\left\{ P_1, P_2 \right\} = \left\{ 46, 20 \right\}$dBm. The user uplink transmission power before power control is $P_u = 20$dBm. We set $\tau_k = 3$dB (or $\tau_k=2$ in linear scale)
for both downlink and uplink in both tiers.

\subsection{Validation of the Optimization of User Association Bias and Spectrum Partition Ratio}

First, we validate the optimization results of the bias factors under spectrum sharing. We then validate the optimal spectrum partition ratio under orthogonal spectrum allocation. We collect the mean user log-utility averaged over multiple snapshots, and also show the mean rate and cell-edge $5$th percentile rate\footnote{Value at the 5\% point of the cumulative density function of the user rate.} for reference. In the simulation, since the user performance in each snapshot is averaged over a finite number of time slots, a small portion of users may experience zero data rate (i.e., outage happens in all subcarriers during all time slots). The log-utility is $-\infty$ for these users, so we only count the mean utility of users with non-zero rate, but also collect the proportion of zero-rate users. The actual mean utility is therefore the combined effect of the mean finite utility and the zero-rate user proportion. The two metrics are shown in the same (left) figure against two separate axes in Figs. \ref{Fig:OptimalBiasDownlink}-\ref{Fig:OptimalPartitionUplink}. The mean rate and cell-edge rate are also plotted in the same (right) figure in Figs. \ref{Fig:OptimalBiasDownlink}-\ref{Fig:OptimalPartitionUplink} for ease of comparison. In all figures in this part of simulation, the ranges of $y$-axis of the mean rate, cell-edge $5$th percentile rate, utility, and proportion of zero-rate users are set to $0.1\sim1.3$Mbps, $0\sim0.3$Mbps, $-2.6\sim-0.2$ (in log(Mbps)), and $1\%\sim10\%$, respectively.

\subsubsection{Optimization of User Association Bias}

\begin{figure}[!t]
\centering
\includegraphics[width=3.5in]{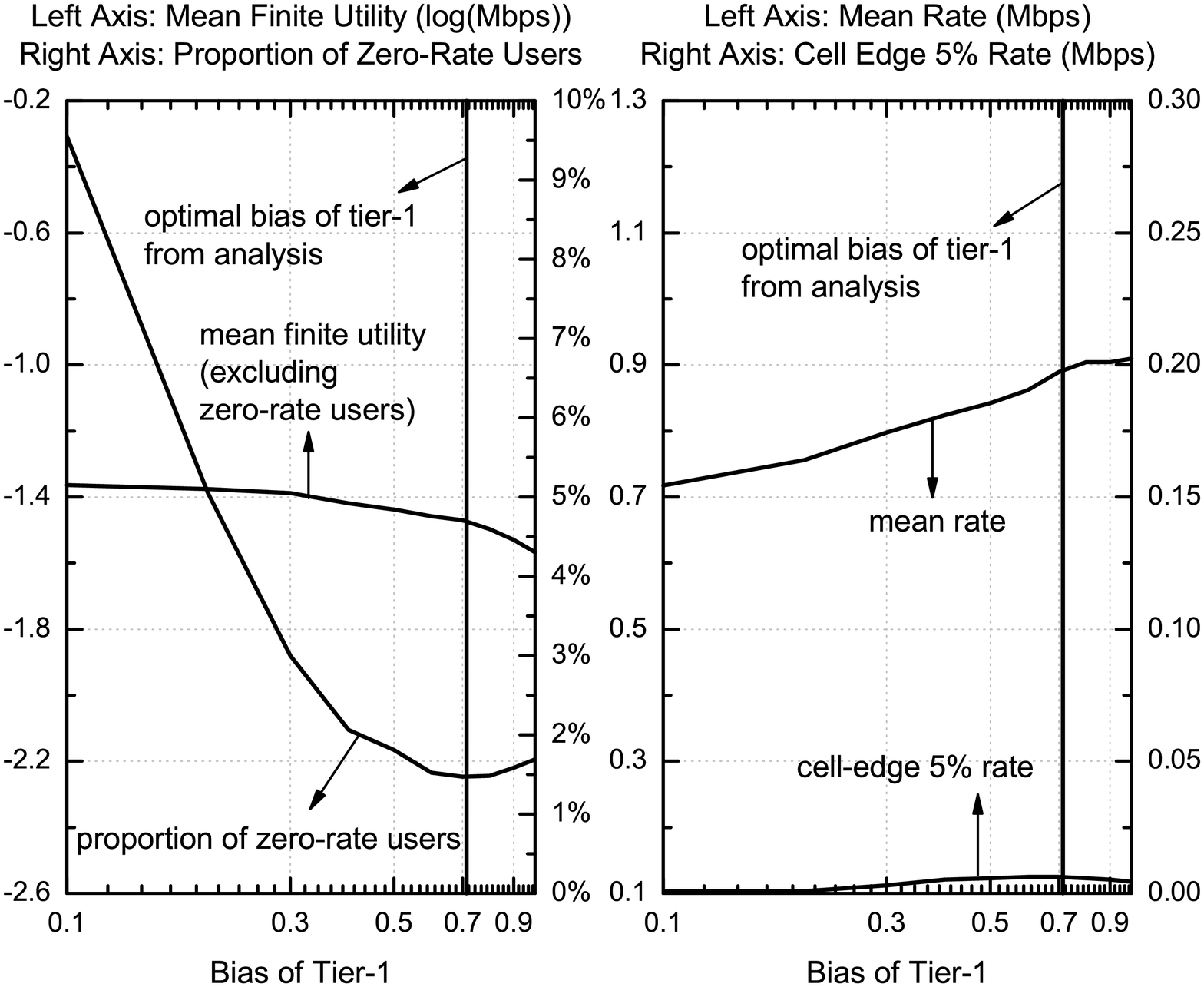}
\caption{Downlink performances vs. bias of tier-1 under spectrum sharing. $\alpha=4$, $\left\{ P_1, P_2 \right\} = \left\{ 46, 20 \right\}$dBm, $\left\{ \lambda_1, \lambda_2 \right\} = \left\{ 0.01, 0.09 \right\} \lambda_u$, $\left\{ \tau_1, \tau_2 \right\} = \left\{ 2, 2 \right\}$.} \label{Fig:OptimalBiasDownlink}
\end{figure}

\begin{figure}[!t]
\centering
\includegraphics[width=3.5in]{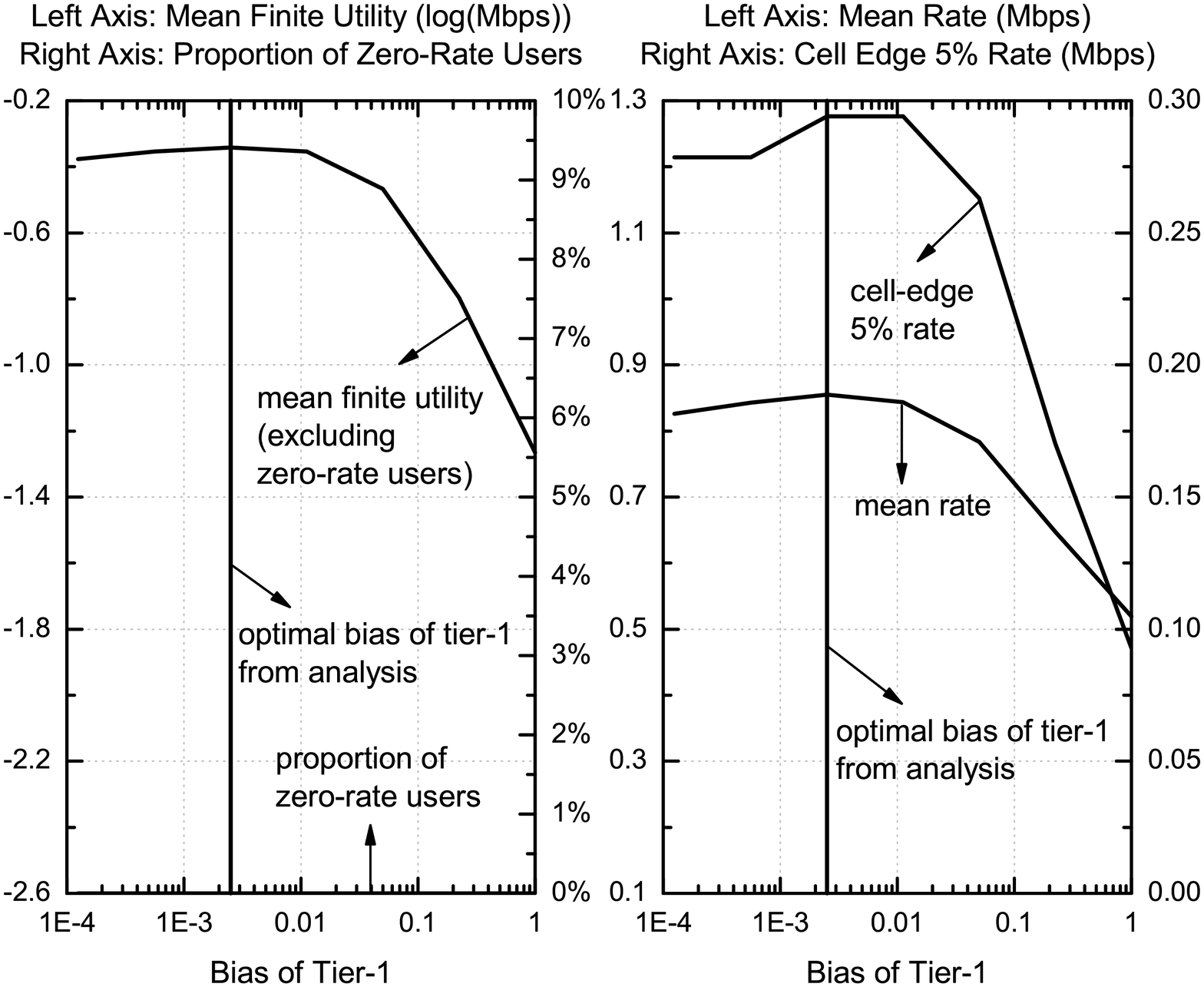}
\caption{Uplink performances vs. bias of tier-1 under spectrum sharing. $\alpha=4$, $\left\{ P_1, P_2 \right\} = \left\{ 46, 20 \right\}$dBm, $P_u = 20$dBm, $\epsilon = 1$, $\left\{ \lambda_1, \lambda_2 \right\} = \left\{ 0.01, 0.09 \right\} \lambda_u$, $\left\{ \tau_1, \tau_2 \right\} = \left\{ 2, 2 \right\}$.} \label{Fig:OptimalBiasUplink}
\end{figure}

Since the optimal bias factor of tier-2 is normalized to $0$dB, we vary the bias $B_1$ of tier-1 and obtain the simulation results in Fig. \ref{Fig:OptimalBiasDownlink} and Fig. \ref{Fig:OptimalBiasUplink} for downlink and uplink, respectively. As observed in Fig. \ref{Fig:OptimalBiasDownlink}, for downlink under spectrum sharing, the mean finite utility and mean rate do not change much within the simulated range. However, a large number of users receive zero rate, and the cell edge rate is very low. This large proportion of zero-rate users dominate the utility computation. We observe that the analytically derived bias factor $B_1^*$ from solving problem (\ref{Eq:OptimalAssociationProblemDownlinkShared}) (marked as vertical lines) achieves nearly the smallest proportion of zero-rate users and the largest cell edge rate.

For uplink under spectrum sharing in Fig. \ref{Fig:OptimalBiasUplink}, as $\tau_1 = \tau_2$ and $\epsilon \alpha=4$, from Theorem 3 we know that the optimal user association is distance based, i.e., the solution to problem (\ref{Eq:OptimalAssociationProblemUplinkShared}) satisfies $B_1 = \frac{P_2}{P_1}B_2 \approx 0.0025$ (with $B_2$ normalized to 1). This analytically optimal value is validated from Fig. \ref{Fig:OptimalBiasUplink} as it indeed simultaneously nearly maximizes the utility, rate and also cell-edge rate. Note that in Fig. \ref{Fig:OptimalBiasUplink} there are no zero-rate users in the simulation, as we assume full power control ($\epsilon=1$) and the path loss of the signal strength is fully compensated.

We also observe that the utility and cell edge rate of the uplink with full power control are significantly greater than those of the downlink under spectrum sharing. This is due in part to the fact that full power control is well suited for our transmission model, in which a fixed target SIR is set in each frequency resource block.

\subsubsection{Optimization of Spectrum Partition Ratio}

\begin{figure}[!t]
\centering
\includegraphics[width=3.5in]{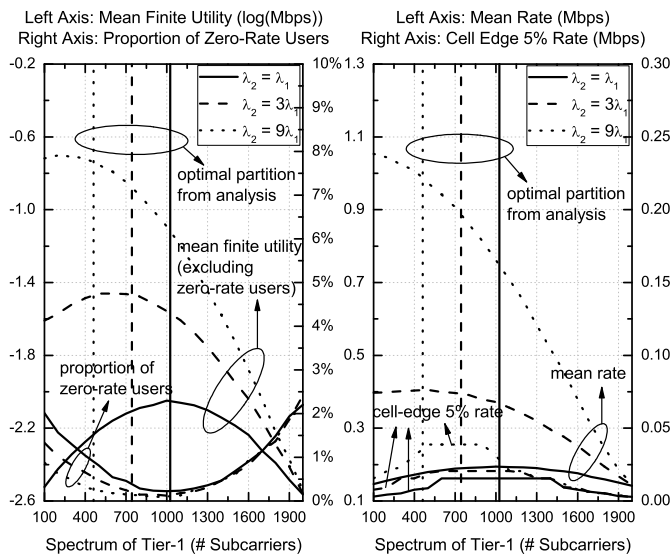}
\caption{Downlink performances vs. inter-tier spectrum partition. The x-axis is the number of subcarriers (out of 2048) allocated to tier-1. The rest of the subcarriers are allocated to tier-2. $\alpha=4$, $\left\{ P_1, P_2 \right\} = \left\{ 46, 20 \right\}$dBm, $\left\{ \lambda_1, \lambda_2 \right\} = \left\{ 0.01, 0.01\sim0.09 \right\} \lambda_u$, $\left\{ \tau_1, \tau_2 \right\} = \left\{ 2, 2 \right\}$.} \label{Fig:OptimalPartitionDownlink}
\end{figure}

\begin{figure}[!t]
\centering
\includegraphics[width=3.5in]{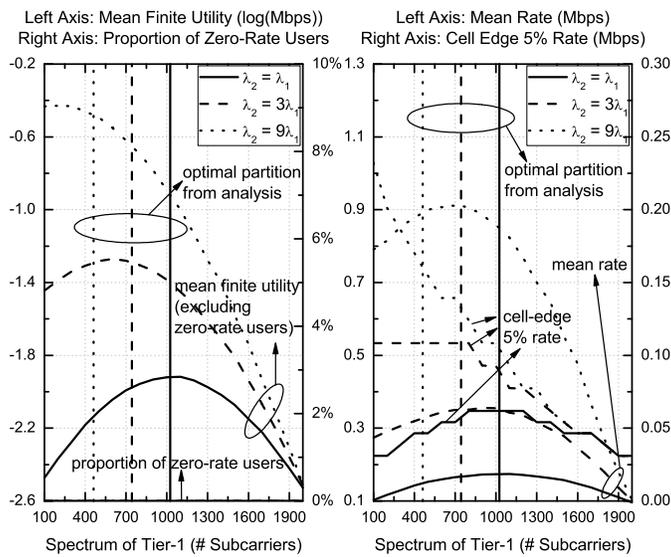}
\caption{Uplink performances vs. inter-tier spectrum partition. The x-axis is the number of subcarriers (out of 2048) allocated to tier-1. The rest of the subcarriers are allocated to tier-2. $\alpha=4$, $\left\{ P_1, P_2 \right\} = \left\{ 46, 20 \right\}$dBm, $P_u = 20$dBm, $\epsilon = 1$, $\left\{ \lambda_1, \lambda_2 \right\} = \left\{ 0.01, 0.01\sim0.09 \right\} \lambda_u$, $\left\{ \tau_1, \tau_2 \right\} = \left\{ 2, 2 \right\}$.} \label{Fig:OptimalPartitionUplink}
\end{figure}

We plot utility and rate against the number of subcarriers (out of a total of $2048$ subcarriers) that are allocated to tier-1 in Fig. \ref{Fig:OptimalPartitionDownlink} and Fig. \ref{Fig:OptimalPartitionUplink} for downlink and uplink, respectively. The remaining subcarriers are allocated to tier-2. The spectrum partition ratio $\eta_1$ (or $\eta_2$) is the ratio of subcarrier number of tier-1 (or tier-2) over $2048$. As discussed in Section IV, the optimal partition ratio is equal to the optimal association probability, hence we set $A_1 = \eta_1$ for each value of $\eta_1$ in the figures, and $B_1$ and $B_2$ can be computed from $A_1$ and $A_2$ using (\ref{Eq:BasA}). The optimal $\eta_1^*$ analytically derived in (\ref{Eq:OptimalAssociationDownlinkOrthogonal}) is also plotted as vertical lines for reference. With $\epsilon=1$ for uplink, by Corollary 1, the optimal spectrum partition of uplink and downlink are the same, which is validated in the figures. The analytically optimal values approximately achieve the highest utility, and strike a balance between maximizing the mean rate and maximizing the cell-edge rate (since the maximal mean rate and maximal cell-edge rate do not result in the same spectrum partition). Note that when both tiers have the same deployment intensity, i.e., $\lambda_1=\lambda_2$, the optimal partition scheme allocates equal spectrum to each tier.

We note that, by comparing the spectrum sharing case in Fig. \ref{Fig:OptimalBiasDownlink} and Fig. \ref{Fig:OptimalBiasUplink}, with the orthogonal spectrum partition case in Fig. \ref{Fig:OptimalPartitionDownlink} and Fig. \ref{Fig:OptimalPartitionUplink} (with $\lambda_2 = 9 \lambda_1$ for both cases), the orthogonal allocation of spectrum can significantly improve the utility and cell-edge rate for downlink. However, sharing of spectrum leads to better utility and cell-edge rate for uplink. This is a consequence of the fact that power control is applied to the uplink, but not to the downlink. As a result, the downlink system is more sensitive to interference, while uplink with power control is less sensitive to interference and thus prefers high utilization of spectrum.

\subsection{Optimal Bias and Spectrum Partition Ratio under Different System Parameters}

We now study the influence of system parameters, such as BS deployment density, BS power and uplink power control factor, on the optimal bias factors and spectrum partition ratios.

\begin{figure}[!t]
\centering
\includegraphics[width=3.5in]{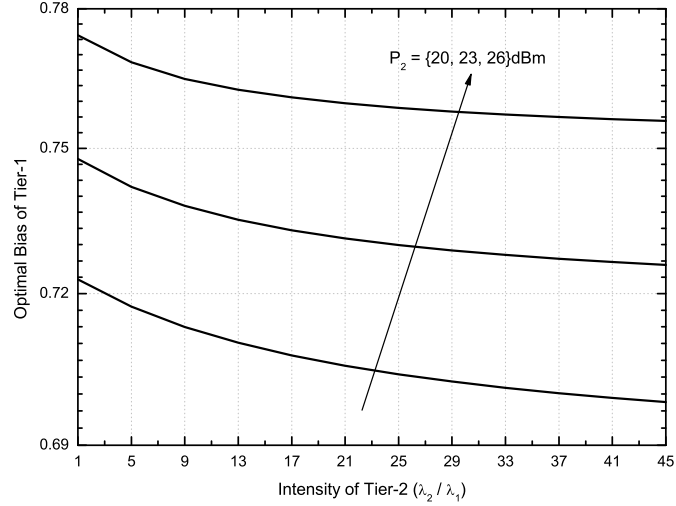}
\caption{Optimal downlink bias of tier-1 vs. intensity of tier-2 under spectrum sharing. $\alpha=4$, $\left\{ P_1, P_2 \right\} = \left\{ 46, 20\sim26 \right\}$dBm, $\left\{ \lambda_1, \lambda_2 \right\} = \left\{ 0.01, 0.01\sim0.45 \right\} \lambda_u$, $\left\{ \tau_1, \tau_2 \right\} = \left\{ 2, 2 \right\}$.} \label{Fig:DownlinkSharedTrend}
\end{figure}

\begin{figure}[!t]
\centering
\includegraphics[width=3.5in]{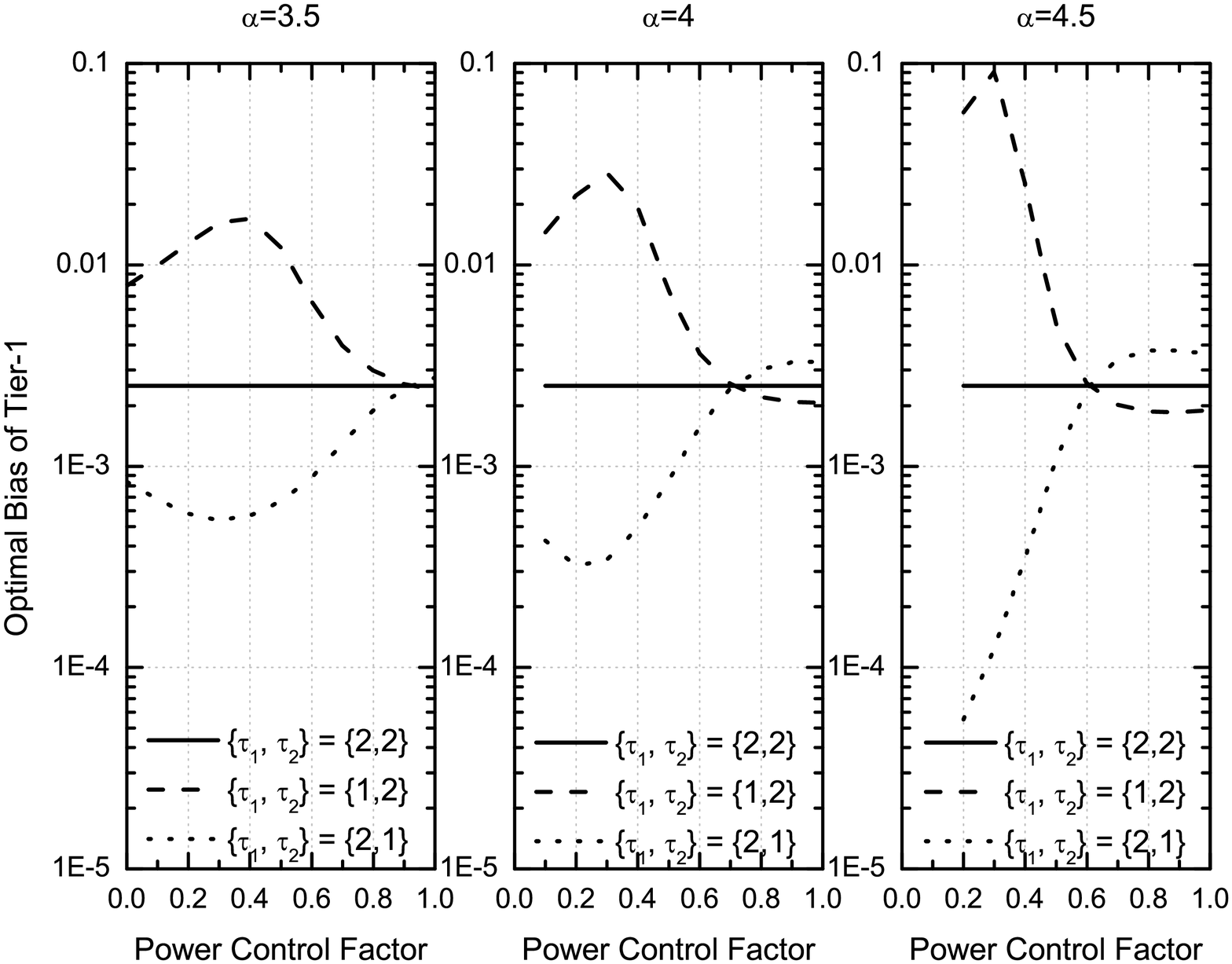}
\caption{Optimal uplink bias of tier-1 vs. power control factor under spectrum sharing. $\alpha=3.5\sim4.5$, $\left\{ P_1, P_2 \right\} = \left\{ 46, 20 \right\}$dBm, $P_u = 20$dBm, $\epsilon = 0\sim1$, $\left\{ \lambda_1, \lambda_2 \right\} = \left\{ 0.01, 0.09 \right\} \lambda_u$.} \label{Fig:UplinkSharedTrend}
\end{figure}

For downlink under spectrum sharing in Fig. \ref{Fig:DownlinkSharedTrend}, as more tier-2 BSs are deployed, more spatial diversity are brought by the new access points, the optimal bias factor of tier-1 BSs hence drops to allow more users to be offloaded to tier-2. With higher tier-2 power, the optimal tier-1 bias factor increases to prevent too many users from being offloaded to tier-2. For uplink under spectrum sharing, under $\alpha=4$, $\epsilon=1$ and $\tau_1=\tau_2$, the optimal user association is distance based and $B_1 = \frac{P_2}{P_1} \approx 0.0025$ is optimal. We vary $\alpha$ and $\epsilon$ in Fig. \ref{Fig:UplinkSharedTrend} to check $\epsilon \alpha$ in the range other than $2$ or $[4, \infty)$ to complement the conclusion in Theorem 3. Note that from (\ref{Eq:PowerControlConstraint}), the feasible regimes of $\epsilon$ for $\alpha=3.5/4/4.5$ are $[0,1]$, $(0,1]$, and $(\frac{1}{9},1]$, respectively. Hence our simulation ranges of $\epsilon$ for cases with $\alpha=4$ and $\alpha=4.5$ in Fig. \ref{Fig:UplinkSharedTrend} are set to $[0.1,1]$ and $[0.2,1]$, respectively. It is observed that, when $\tau_1 = \tau_2$, for all values of $\epsilon \alpha$, including $\epsilon \alpha < 2$ and $2 < \epsilon \alpha < 4$ where the objective function is neither convex nor concave, the optimal bias that maximizes (\ref{Eq:OptimalAssociationProblemUplinkShared}) is still equal to $0.0025$, hence is equivalent to distance-based association. This numerical observation extends Theorem 3 for all feasible regimes of $\epsilon \alpha$. For the case where $\tau_1 \neq \tau_2$ in Fig. \ref{Fig:UplinkSharedTrend}, the optimal association is close to be distance-based at about $\epsilon > 0.6$.

\begin{figure}[!t]
\centering
\includegraphics[width=3.5in]{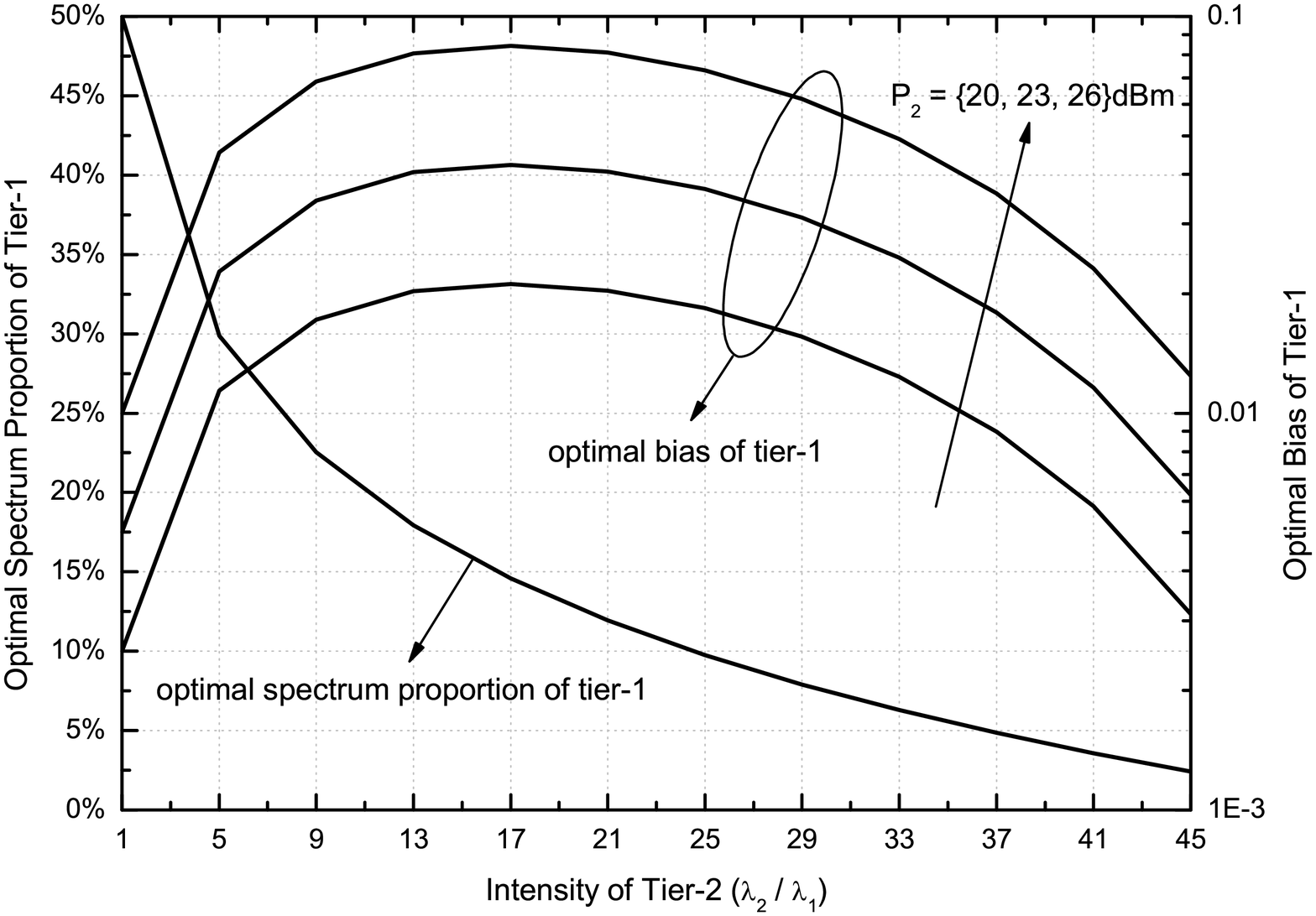}
\caption{Optimal downlink spectrum proportion and bias of tier-1 vs. intensity of tier-2. $\alpha=4$, $\left\{ P_1, P_2 \right\} = \left\{ 46, 20\sim26 \right\}$dBm, $\left\{ \lambda_1, \lambda_2 \right\} = \left\{ 0.01, 0.01\sim0.45 \right\} \lambda_u$, $\left\{ \tau_1, \tau_2 \right\} = \left\{ 2, 2 \right\}$.} \label{Fig:DownlinkOrthogonalTrend}
\end{figure}

\begin{figure}[!t]
\centering
\includegraphics[width=3.5in]{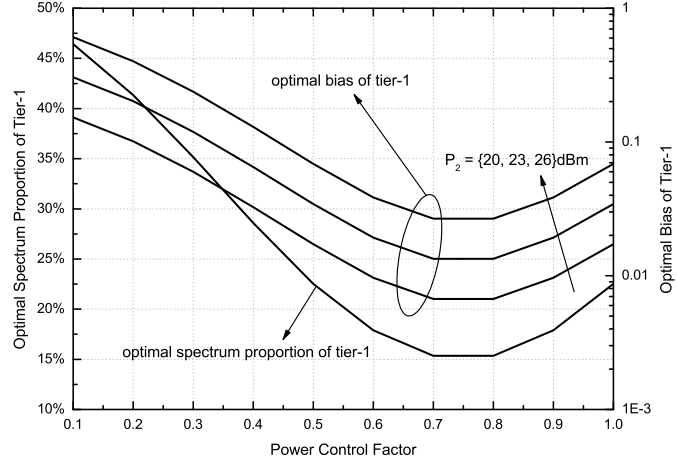}
\caption{Optimal uplink spectrum proportion and bias of tier-1 vs. power control factor. $\alpha=4$, $\left\{ P_1, P_2 \right\} = \left\{ 46, 20\sim26 \right\}$dBm, $P_u = 20$dBm, $\epsilon = 0\sim1$, $\left\{ \lambda_1, \lambda_2 \right\} = \left\{ 0.01, 0.09 \right\} \lambda_u$, $\left\{ \tau_1, \tau_2 \right\} = \left\{ 2, 2 \right\}$.} \label{Fig:UplinkOrthogonalTrend}
\end{figure}

Fig. \ref{Fig:DownlinkOrthogonalTrend} shows that, for downlink under orthogonal spectrum partition, with more tier-2 BSs deployed in the network, the optimal proportion of spectrum allocated to tier-1 decreases. The optimal bias corresponding to the optimal spectrum allocation first increases then decreases. For uplink with full power control and under orthogonal spectrum partition, by Corollary 1, the behavior of the optimal spectrum allocation with respect to the BS deployment intensity is similar to that of downlink under orthogonal spectrum partition. Thus, in Fig. \ref{Fig:UplinkOrthogonalTrend}, we instead focus on the optimal spectrum allocation and bias as a function of the power control factor $\epsilon$. As $\epsilon$ increases, the optimal bias and spectrum proportion of tier-1 first decrease then increase. Intuitively, users associated with tier-1 are farther from their BSs, and with larger power control factor they transmit at higher power and cause stronger interference to users in tier-2. Consequently, the bias and spectrum allocation for tier-1 should decrease as $\epsilon$ increases in order to mitigate interference. Once the power control factor $\epsilon$ is large enough, the improved signal quality of tier-1 users compensates the increased interference. Hence the bias and spectrum allocation of tier-1 would eventually increase again.

\section{Conclusion}

This paper studies the joint user association and spectrum allocation for multi-tier HetNets in the interference-limited regime. We model BSs and users in the network as spatial point processes, and analytically derive a closed-form approximation of the mean proportionally fair utility of the system based on the coverage rate using stochastic geometry for both downlink and uplink. A optimization framework is formulated based on this network utility function. Our solution reveals the downlink-uplink decoupling in user association, and the distance-based association in uplink under a certain condition. Further, when orthogonal spectrum partition is assumed, the spectrum allocated to each tier should match the users associated with that tier. Simulation results verify the accuracy of the analytical results and illustrate the usefulness of  stochastic geometry in the optimization of HetNets.

\section*{Appendix A}

Here we extend the derivation in \cite{SeungMinYu/2013/Downlink} from the single-tier case to the multi-tier case, and present the result for completeness. First, for single-tier networks, the PDF of the size of the normalized Voronoi cell is approximated with a two-parameter gamma function \cite{Ferenc/2007/Size}
\begin{equation} \label{Eq:SizePDF}
f_S (x) = \frac{3.5^{3.5}}{\Gamma(3.5)} x^{2.5} e^{-3.5x}.
\end{equation}
The PDF of the size of the normalized Voronoi cell, conditioned on that the typical user is associated with that cell (denoted as $\Lambda$), is derived from (\ref{Eq:SizePDF}) as \cite{SeungMinYu/2013/Downlink}
\begin{equation} \label{Eq:TaggedSizePDF}
f_{S|\Lambda} (x) = \frac{3.5^{4.5}}{\Gamma(4.5)} x^{3.5} e^{-3.5x} \overset{\Gamma(t+1) = t \Gamma(t)}{=} \frac{3.5^{3.5}}{\Gamma(3.5)} x^{3.5} e^{-3.5x}.
\end{equation}

With a given cell size, the number of users associated with a BS follows a Poisson distribution. Since the multi-tier cell topology forms a multiplicatively weighted Voronoi tessellation, using the area approximation in \cite{SarabjotSingh/2013/Offloading,SarabjotSingh/2014/Joint}, the probability mass function (PMF) of the number of users associated with a tier-$k$ BS is derived using (\ref{Eq:SizePDF}) as
%\begin{align} \label{Eq:UserNumPMF}
%\mathbb{P} (N_k = n) &= \int_0^\infty \frac{ \left( \frac{ A_k \lambda_u }{ \lambda_k } x \right)^n }{ n! } e^{ - \frac{ A_k \lambda_u }{ \lambda_k } x } f_S (x) \text{d}x \nonumber \\
%&= \frac{ 3.5^{3.5} \Gamma(n+3.5) \left( A_k \lambda_u / \lambda_k \right)^n }{ \Gamma(3.5) n! \left( A_k \lambda_u / \lambda_k + 3.5 \right)^{n+3.5} }
%\end{align}
\begin{equation} \label{Eq:UserNumPMF}
\mathbb{P} (N_k = n) = \int_0^\infty \frac{ \left( \frac{ A_k \lambda_u }{ \lambda_k } x \right)^n }{ n! } e^{ - \frac{ A_k \lambda_u }{ \lambda_k } x } f_S (x) \text{d}x = \frac{ 3.5^{3.5} \Gamma(n+3.5) \left( A_k \lambda_u / \lambda_k \right)^n }{ \Gamma(3.5) n! \left( A_k \lambda_u / \lambda_k + 3.5 \right)^{n+3.5} }
\end{equation}
and the PMF of the number of other users (apart from the typical user) of a tier-$k$ BS, conditioned on the typical user being associated with that BS, is derived in a similar way using (\ref{Eq:TaggedSizePDF}) as
\begin{equation} \label{Eq:TaggedUserNumPMF}
\mathbb{P} (\tilde{N}_k = n) = \frac{ 3.5^{3.5} \Gamma(n+4.5) \left( A_k \lambda_u / \lambda_k \right)^n }{ \Gamma(3.5) n! \left( A_k \lambda_u / \lambda_k + 3.5 \right)^{n+4.5} }.
\end{equation}

Hence the mean proportion of spectrum allocated to the typical user associated with BS tier-$k$ is
\begin{align}
\mathbb{E} \left( \frac{1}{\tilde{N}_k + 1} \right) &= \sum_{n=0}^\infty \frac{1}{n+1} \mathbb{P} \left( \tilde{N}_k = n \right) \nonumber \\
&\overset{(a)}{=} \frac{ \lambda_k }{ A_k \lambda_u } \sum_{i=1}^\infty \frac{ 3.5^{3.5} \Gamma(i+3.5) \left( A_k \lambda_u / \lambda_k \right)^i }{ \Gamma(3.5) i! \left( A_k \lambda_u / \lambda_k + 3.5 \right)^{i+3.5} } \nonumber \\
&= \frac{ \lambda_k }{ A_k \lambda_u } \left[ \sum_{i=0}^\infty \mathbb{P} \left( N_k = i \right) - \mathbb{P} \left( N_k = 0 \right) \right] \nonumber \\
&= \frac{ \lambda_k }{ A_k \lambda_u } \left[ 1 - \left( 1 + \frac{ A_k \lambda_u }{ 3.5 \lambda_k } \right)^{-3.5} \right] \nonumber \\
&\lesssim \frac{ \lambda_k }{ A_k \lambda_u },
\end{align}
where in (a) the change of variables $i=n+1$ is used. The last upper bound is tight for wireless systems with a large number of users per base-station, i.e., $\frac{A_k \lambda_u}{\lambda_k} \gg 1$, where the term $\left( 1 + \frac{ A_k \lambda_u }{ 3.5 \lambda_k } \right)^{-3.5}$ is negligible. Since $\beta_k = \frac{W_k}{\tilde{N}_k + 1}$ and $W_k$ is fixed, we have the result in (\ref{Eq:PerUserResource}).

\section*{Appendix B}

We separate the objective in (\ref{Eq:OptimalAssociationProblemUplinkShared}) into three terms, and form three subproblems:
\begin{subequations} \label{Eq:OptimalAssociationSubProblemUplinkShared}
\begin{align}
&\max_{A_k, \forall k} \sum_{k=1}^K A_k \log \left[ \frac{W \lambda_k \log (1 + \tau_k)}{A_k \lambda_u} \right], \label{Eq:OptimalAssociationSubProblemUplinkShared1} \\
&\min_{A_k, \forall k} \sum_{i=1}^K \tau_i \lambda_i^{\epsilon \alpha/2 - 1} A_i^{2 - \epsilon \alpha/2}, \label{Eq:OptimalAssociationSubProblemUplinkShared2} \\
&\min_{A_k, \forall k} \sum_{j=1}^K \frac{A_j^{\epsilon \alpha/2}}{\lambda_j^{\epsilon \alpha/2 - 1}}, \label{Eq:OptimalAssociationSubProblemUplinkShared3}
\end{align}
\end{subequations}
subject to the constraints (\ref{Eq:OptimalAssociationProblemDownlinkShared2}) and (\ref{Eq:OptimalAssociationProblemDownlinkShared3}).

The objective of subproblem (\ref{Eq:OptimalAssociationSubProblemUplinkShared1}) is concave. First, forming Lagrangian with respect to its constraint (\ref{Eq:OptimalAssociationProblemDownlinkShared2}) and taking the first order derivative to $A_k$, after some simplifications we have the solution to subproblem (\ref{Eq:OptimalAssociationSubProblemUplinkShared1}) as $A_k^{(1)} = \log(1+\tau_k) \lambda_k \omega$. Then, by applying $\sum_{k=1}^K A_k^{(1)} = 1$ to obtain the dual variable $\omega$, we have the following optimal user association probability for subproblem (\ref{Eq:OptimalAssociationSubProblemUplinkShared1}):
\begin{equation}
A_k^{(1)} = \frac{ \log(1+\tau_k) \lambda_k }{ \sum_{j=1}^K \log(1+\tau_j) \lambda_j }.
\end{equation}

If $\epsilon \alpha \ge 4$, the objectives of both subproblems (\ref{Eq:OptimalAssociationSubProblemUplinkShared2}) and (\ref{Eq:OptimalAssociationSubProblemUplinkShared3}) are convex, and their solutions can be obtained in a similar way as for subproblem (\ref{Eq:OptimalAssociationSubProblemUplinkShared1})
\begin{align}
A_k^{(2)} &= \frac{ \tau_k^{2 / (\epsilon\alpha-2)} \lambda_k }{ \sum_{j=1}^K \tau_j^{2 / (\epsilon\alpha-2)} \lambda_j }, \\
A_k^{(3)} &= \frac{ \lambda_k }{ \sum_{j=1}^K \lambda_j }.
\end{align}
If $\tau_k = \tau_j$, $\forall j \neq k$, the three solutions $A_k^{(1)} = A_k^{(2)} = A_k^{(3)}$, and are equal to
\begin{equation} \label{Eq:OptimalAssociationUplinkSharedSpecial}
A_k^* = \frac{\lambda_k}{\sum_{j=1}^K \lambda_j},
\end{equation}
which is also the optimal solution to the original problem (\ref{Eq:OptimalAssociationProblemUplinkShared}).

If $\epsilon \alpha = 2$, the objective of subproblem (\ref{Eq:OptimalAssociationSubProblemUplinkShared2}) becomes $\sum_i \tau_i A_i = \tau$ since $\tau_i = \tau$, $\forall i$, and the objective of subproblem (\ref{Eq:OptimalAssociationSubProblemUplinkShared3}) becomes $\sum_j A_j = 1$, both of which are constants. Solution to problem (\ref{Eq:OptimalAssociationProblemUplinkShared}) is hence $A_k^{(1)}$ determined from subproblem (\ref{Eq:OptimalAssociationSubProblemUplinkShared1}), which equals (\ref{Eq:OptimalAssociationUplinkSharedSpecial}) if $\tau_k = \tau_j$, $\forall j \neq k$.

The corresponding bias factors for association probability in (\ref{Eq:OptimalAssociationUplinkSharedSpecial}) can be obtained via the transformation (\ref{Eq:BasA}), and are found to satisfy $P_k B_k = P_j B_j$, $\forall k \neq j$. From (\ref{Eq:AssociationRule}), we know that this is equivalent to the distance-based user association.

If $\epsilon \alpha < 2$, the objective of (\ref{Eq:OptimalAssociationSubProblemUplinkShared2}) is convex and the objective of (\ref{Eq:OptimalAssociationSubProblemUplinkShared3}) is concave; if $2 < \epsilon \alpha < 4$, the objective of (\ref{Eq:OptimalAssociationSubProblemUplinkShared2}) is concave and the objective of (\ref{Eq:OptimalAssociationSubProblemUplinkShared3}) is convex. In these cases, the original problem (\ref{Eq:OptimalAssociationProblemUplinkShared}) is neither convex nor concave.

\balance

%\bibliographystyle{IEEEtran}
% Generated by IEEEtran.bst, version: 1.13 (2008/09/30)

%\bibliography{IEEEabrv,mybibfile}

%\begin{IEEEbiography}[{\includegraphics[trim = 0mm 10mm 0mm 0mm,clip,width=25mm,keepaspectratio]{Yicheng_Lin.eps}}]{Yicheng Lin}
%(S¡¯09) received the ... degree(s)...
%\end{IEEEbiography}

%\begin{IEEEbiography}[{\includegraphics[trim = 25mm 0mm 30.5mm 0mm,clip,width=25mm,keepaspectratio]{Wei_Yu.eps}}]{Wei Yu}
%(S¡¯97-M¡¯02-SM¡¯08) received the ... degree(s)...
%\end{IEEEbiography}

\end{document}